\documentclass[a4paper,11pt]{article}
\pdfoutput=1
\usepackage{geometry}
\usepackage{a4wide,slashbox}
\usepackage{graphicx}
\usepackage{epsf}
\usepackage{amsmath}
\usepackage[normalem]{ulem}
\usepackage{amssymb}

\usepackage{cite}
\usepackage{multirow,tabularx}
\usepackage{appendix}
\usepackage{tikz}
\usepackage{graphicx,amsmath,amsfonts,amssymb,amsthm,euscript,braket,xcolor}
\newcommand{\be}{\begin{equation}}
\newcommand{\ee}{\end{equation}}

\newcommand{\Rmnum}[1]{\expandafter\@slowromancap\romannumeral #1@}
\newcommand{\bea}{\begin{eqnarray}}
\newcommand{\eea}{\end{eqnarray}}

\numberwithin{equation}{section}

\newcommand*\circled[1]{\tikz[baseline=(char.base)]{
            \node[shape=circle,draw,inner sep=2pt] (char) {#1};}}

\usepackage{subfigure}
\begin{document}

\title{\bf Exact topological charged hairy black holes in AdS Space in $D$-dimensions}

\author{\textbf{Subhash Mahapatra}\thanks{mahapatrasub@nitrkl.ac.in}, \textbf{Supragyan Priyadarshinee}\thanks{518ph1009@nitrkl.ac.in},
 \textbf{Gosala Narasimha Reddy}\thanks{415ph5049@nitrkl.ac.in},
 \\ \textbf{Bhaskar Shukla}\thanks{519ph1003@nitrkl.ac.in}
 \\\\
 \textit{{\small Department of Physics and Astronomy, National Institute of Technology Rourkela, Rourkela - 769008, India}}
}
\date{}


\maketitle
\abstract{We present a new family of topological charged hairy black hole solutions in asymptotically AdS space in $D$-dimensions. We solve the coupled Einstein-Maxwell-Scalar gravity system and obtain exact charged hairy black hole solutions with planar, spherical and hyperbolic horizon topologies. The scalar field is regular everywhere. We discuss the thermodynamics of the hairy black hole and find drastic changes in its thermodynamic structure due to the scalar field. For the case of planar and spherical horizons, we find charged hairy/RN-AdS black hole phase transition, with thermodynamically preferred and stable charged hairy phases at low temperature. For the case of hyperbolic horizon, no such transition occurs and RN-AdS black holes are always thermodynamically favoured.}

\section{Introduction}
Black holes are very interesting objects of general relativity (GR) where both quantum physics and strong gravity are believed to coexist. These are the simplest and yet the most mysterious solutions of GR that have been an intriguing subject of research for many decades and still are far from being fully understood.  Classical black holes in GR are believed to follow the famous no-hair theorem, which simply states that spherically symmetric black hole in the asymptotically flat background are completely characterised by its mass, charge and angular momentum \cite{Ruffini}. In other words, black holes in asymptotically flat space do not support static scalar field hair outside their horizon. The main reason for this belief is partly based on the absorbing nature of the horizon.  Although, the initial black hole no-hair theorem of \cite{Ruffini} had received support from many other works, see for example \cite{Bekenstein:1971hc,Israel:1967wq,Israel:1967za,Wald:1971iw,Carter:1971zc,Robinson:1975bv,Mazur:1982db,Mazur:1984wz,Teitelboim:1972qx,Chase}, however it is not a theorem in the usual mathematical sense and by now many counterexamples exist, see \cite{Volkov:1990sva,Bizon:1990sr,Kuenzle:1990is,Straumann:1990as,Zhou:1991nu,Bizon:1991hw,Bizon:1991nt,Volkov:1995np,Brodbeck:1994vu,Garfinkle:1990qj,Herdeiro:2014goa,Berti:2013gfa,Greene:1992fw,Lavrelashvili:1992ia,
Torii:1993vm} and discussion below. The examination of the scalar-gravity system and no-hair theorem is not just a theoretical curiosity.  From the astronomical observational point of view, high precision measurements of the super-massive black holes may provide experimental verification of the no-hair theorem \cite{Sadeghian:2011ub}.
\\
\\
There are mainly two major requirements for a stable scalar hairy black hole solution. First, the scalar field should be regular at the horizon, and second, it should fall sufficiently fast at the asymptotic boundary. Apart from these requirements, the hairy black hole solution should not also lead to any additional singularity outside the horizon and it should be stable under perturbations to be considered physically expectable. To the best of our knowledge, initial attempts to investigate minimally coupled scalar-gravity system were done in \cite{Bocharova,Bekenstein:1974sf,Bekenstein:1975ts} for the asymptotic flat spaces. However, the constructed scalar hair solution turned out to be un-physical as the scalar field diverged on the horizon and stability analysis further confirmed its unstable nature \cite{Bronnikov:1978mx}. A straightforward proof of the no-scalar hair theorem for the spherically symmetric asymptotic flat spacetime was provided in \cite{Bekenstein:1971hc,Bekenstein:1995un,Sudarsky:1995zg}, see also \cite{Heusler:1992ss}. For a recent discussion and review on the issue of scalar hair in asymptotic flat spaces, see \cite{Herdeiro:2015waa,Hertog:2006rr,Astefanesei:2019mds}. It was soon realised that the introduction of the cosmological constant may pave the way for a regular scalar field configuration near the horizon, as the cosmological constant may act as an effective potential outside the horizon which may stabilise the scalar field. Hairy black hole solution for the minimally coupled scalar-gravity system in the asymptotic dS space was obtained in \cite{Zloshchastiev:2004ny}, however, it turned out to be unstable \cite{Torii:1998ir}. In the case of asymptotically AdS space, stable hairy black hole solutions with various horizon topologies were obtained and analysed in \cite{Torii:1998ir,Torii:2001pg,Winstanley:2002jt,Martinez:2004nb,Martinez:2005di,Martinez:2006an,Hertog:2004dr,Henneaux:2004zi,Henneaux:2006hk,Amsel:2006uf}. In recent years, countless works addressing various physical aspects of the hairy black holes in different asymptotic spaces have appeared, for a necessarily biased selection see \cite{Kolyvaris:2009pc,Dias:2011at,Dias:2011tj,Bhattacharyya:2010yg,Basu:2010uz,Anabalon:2012ta,Anabalon:2012ih,Anabalon:2012tu,Kleihaus:2013tba,Kolyvaris:2011fk,Kolyvaris:2013zfa,Gonzalez:2013aca,Anabalon:2009qt,Charmousis:2009cm}.
\\
\\
The interplay between scalar field and black hole geometry has attracted much attention of late, especially in the context of AdS spaces. One of the main reasons for this is the discovery of gauge/gravity duality \cite{Maldacena:1997re,Witten:1998qj,Gubser:1998bc}.  The gauge/gravity duality provides an excellent technical tool for investigating strongly coupled quantum theories using classical gravity in AdS spaces, and has found applications in wide range of areas ranging from QCD to condensed matter. The duality also provides an appropriate way to introduce external parameters such as temperature, chemical potential, magnetic field etc in the boundary theory, which can then be used to model realistic strongly coupled systems via holography. Using this duality, black holes in AdS space have been extensively used to address interesting features of strongly coupled systems such as thermalization, transport coefficients, confinement/deconfinement physics etc \cite{Kovtun:2004de,Balasubramanian:2010ce,Dey:2015poa,Witten:1998zw}. Similarly, hairy black holes in AdS space have also appeared abundantly in applied holographic theories, with the most noticeable application occurring in the holographic studies of superconductors (and other condensed matter phenomena). The whole idea of holographic superconductors is based on the dynamical study of a charged scalar field outside the horizon \cite{Gubser:2008px,Hartnoll:2008vx,Hartnoll:2008kx}. In particular, how the charged scalar field develops an effective negative mass and condenses outside the RN-AdS horizon, thereby making the hairy RN-AdS solution thermodynamically more favourable at low temperatures, is the essence of holographic superconductors. 
\\
\\
Black holes in AdS spaces are also quite different from their dS or flat spaces counterparts. In asymptotically dS or flat spaces, the horizon topology of a four-dimensional black hole must be a round sphere $\mathbb{S}^2$ whereas the horizon topology can be a planar $\mathbb{R}^2$, sphere $\mathbb{S}^2$ or hyperbolic $\mathbb{H}^2$ in AdS spaces \cite{Friedman:1993ty,Birmingham:1998nr,Brill:1997mf,Mann:1996gj,Lemos:1994xp,Vanzo:1997gw,Cai:1996eg}. Correspondingly, the associated thermodynamic properties and phase transitions are much more interesting in AdS black holes. In particular, spherical Schwarzschild AdS black hole are not only stable with its surrounding but can undergo Hawking/Page thermal-AdS/black hole phase transition \cite{Hawking:1982dh}. Similarly, spherical charged RN-AdS black holes can exhibit Van der Waals type small/large black hole phase transitions and display a rather rich phase structure \cite{Chamblin:1999tk,Chamblin:1999hg,Dey:2015ytd,Mahapatra:2016dae,Cvetic:1999ne}. For a discussion on the thermodynamics with the hyperbolic horizon, see \cite{Brill:1997mf,Cai:2004pz,Sheykhi:2007wg}. Therefore, one can expect that introduction of the scalar hair might modify the thermodynamic properties of the AdS black holes in a non-trivial way and are worth investigating.
\\
\\
From the above discussion, it is clear that it is important to investigate the matter scalar field outside the black hole and, in particular, to construct stable scalar hairy black hole configurations. Indeed, one of the primary aims and interests in GR still now is to look for the exact solutions of Einstein field equations. It is fair to say that among these exact solutions, black holes occupy an important position as they provide a successful framework to unify thermodynamics, gravitational and quantum theories. In addition, scalar fields also play a central role in cosmology and appear naturally in UV complete theory of quantum gravity such as string theory. It is then of utmost importance to understand generic properties of gravity theories coupled to the scalar fields, particularly the role played by the scalar fields to black hole physics.
\\
\\
In this work, our aim is to investigate thermodynamic properties of a family of exact charged/uncharged asymptotically AdS hairy black holes, with the goal of shedding light on their thermodynamic stability. For this purpose, we consider the  Einstein-Maxwell-Scalar gravity system. Using the potential reconstruction method \cite{He:2013qq,Yang:2015aia,Dudal:2017max,Dudal:2018ztm,Mahapatra:2018gig,Mahapatra:2019uql,Bohra:2019ebj,Cai:2012xh,Alanen:2009xs,Arefeva:2018hyo,Arefeva:2018cli},  we solve the Einstein, Maxwell and scalar equations of motion analytically in all spacetime dimensions in terms of an arbitrary scale function $A(z)$ and obtain an infinite family of analytic black hole solutions with planar, spherical and hyperbolic horizon topologies. We choose a particular form of the scale function $A(z)=-a z^n$, which allows us to introduce two parameters $a$ and $n$ in our gravity model. Different values of $a$ and $n$ therefore correspond to different black hole solutions. The gravity solution displays the following admirable features, (i) the scalar field is regular at the horizon and falls off at asymptotic boundary, (ii) the Kretschmann scalar is finite everywhere outside the horizon, suggesting no additional singularity in the geometry and (iii) the scalar potential is bounded from above from its UV boundary value and reduces to the negative cosmological constant at the asymptotic boundary. We then study the thermodynamics of the constructed charged hairy topological black holes and find that the charged planar and spherical hairy black holes are not only thermodynamically stable but also preferable than the RN-AdS black hole at low temperatures, whereas RN-AdS black hole is preferable at high temperatures. However, uncharged hairy black holes always have higher free energy than the Schwarzschild AdS black hole. With the hyperbolic horizon, both charged and uncharged hairy black holes have higher free energy than the RN-AdS black hole.
\\
\\
The paper is organised as follows. In the next section, we introduce the Einstein-Maxwell-Scalar gravity model and present its analytic solution corresponding to the topological charged hairy black hole in all spacetime dimensions. In section 3, we study the thermodynamics of hairy black hole solutions with different horizon topologies. Finally, in section 4, we summarize the main results and point out the future directions.

\section{Exact charged hairy black hole solutions}
To construct and study charged hairy black hole solutions in $D$ spacetime dimensions, we begin with the following Einstein-Maxwell-scalar action,
\begin{eqnarray}
S_{EM} =  -\frac{1}{16 \pi G_D} \int \mathrm{d^D}x \ \sqrt{-g}  \ \left[R - \frac{f(\phi)}{4}F_{MN}F^{MN} -\frac{1}{2}\partial_{M}\phi \partial^{M}\phi -V(\phi)\right]\,.
\label{actionEF}
\end{eqnarray}
where $G_D$ is the $D$-dimensional Newton constant\footnote{$G_D$ will be set to one in the numerical calculations.}, $F_{MN}$ is the field strength tensor of $U(1)$ gauge field $A_M$,  $\phi$ is the scalar field and $f(\phi)$ represents the coupling between $U(1)$ gauge and scalar field. $V(\phi)$ is the potential of the field $\phi$, whose explicit form will depend on the scale function $A(z)$ (see below).\\

The variation of the action (\ref{actionEF}) gives the following Einstein, Maxwell and scalar equations,
\begin{eqnarray}
& & R_{MN}- \frac{1}{2}g_{MN} R + \frac{f(\phi)}{4}\biggl(\frac{g_{MN}}{2} F^2 - 2 F_{MP}F_{N}^{\ P}\biggr)    \nonumber \\
& &  + \frac{1}{2} \biggl(\frac{g_{MN}}{2} \partial_{P}\phi \partial^{P}\phi -\partial_{M}\phi \partial_{N}\phi  + g_{MN} V(\phi)  \biggr)  =0 \,.
\label{EinsteinEE}
\end{eqnarray}
\begin{eqnarray}
& & \nabla_{M} \biggl[ f(\phi) F^{MN}  \biggr] = 0\,.
\label{MaxwellEE}
\end{eqnarray}
\begin{eqnarray}
& & \frac{1}{\sqrt{-g}}\partial_{M} \biggl[ \sqrt{-g}  \partial^{M}\phi \biggr] - \frac{F^2}{4} \frac{\partial f(\phi)}{\partial \phi} - \frac{\partial V(\phi)}{\partial \phi} = 0 \,.
\label{dilatonEE}
\end{eqnarray}
Since, in this work we are interested in constructing the charged hairy black hole solution with various horizon topologies, we consider the following Ans\"atze for the metric $g_{MN}$, field strength tensor $F_{MN}$ and scalar field $\phi$,
\begin{eqnarray}
& & ds^2=\frac{L^2 e^{2A(z)}}{z^2}\biggl[-g(z)dt^2 + \frac{dz^2}{g(z)} + d\Omega_{\kappa, D-2}^2 \biggr]\,, \nonumber \\
& & \phi=\phi(z), \ \ A_{M}=A_{t}(z)\delta_{M}^{t}\,.
\label{ansatz}
\end{eqnarray}
where $A(z)$ is the scale factor, $g(z)$ is the blackening function and $L$ is the AdS length scale, which will be set to one later on in numerical calculations. As usual, $z$ is the radial coordinate and it runs from $z=0$ (asymptotic boundary) to $z=z_h$ (horizon radius), or to $z=\infty$ for thermal AdS (without horizon). The parameter $\kappa$ indicates the curvature of the $(D-2)$-dimensional metric $d\Omega_{\kappa, D-2}^2$, and it can take three different values, $\{-1,0,+1 \}$, corresponding to hyperbolic, planar, and spherical horizon topologies, respectively.

\[
    d\Omega_{\kappa, D-2}^2=
\begin{cases}
    dx_1^2 + \sum\limits_{i=2}^{D-2}  \prod\limits_{j=1}^{i-1} \sin^2x_j dx_i^2, &  \kappa=1 \\
    \sum\limits_{i=1}^{D-2} dx_i^2, &  \kappa=0 \\
   dx_1^2 + \sinh^2x_1dx_2^2 + \sinh^2x_1 \sum\limits_{i=3}^{D-2}  \prod\limits_{j=2}^{i-1} \sin^2x_j dx_i^2, &  \kappa=-1
\end{cases}
\]

Plugging the above Ans\"atze into eq.~(\ref{EinsteinEE}), we get three Einstein equations of motion,
\begin{eqnarray}
tt: \ A''(z) + A'(z) \left(\frac{3-D}{z}+\frac{g'(z)}{2g(z)} +\frac{(D-3)}{2}A'(z) \right) +\frac{e^{-2A(z)}z^2f(z)A_{t}'(z)^2}{4(D-2)L^2g(z)} \nonumber \\
-\frac{g'(z)}{2zg(z)} +\frac{e^{2A(z)}L^2V(z)}{2(D-2)z^2g(z)} + \frac{(D-1)}{2 z^2}+\frac{\phi'(z)^2}{4(D-2)}-\frac{(D-3)\kappa}{2g(z)} =0\,.
\label{Einsteintt}
\end{eqnarray}
\begin{eqnarray}
zz: \ g'(z) \left(A'(z)-\frac{1}{z}\right)+g(z)\left(\frac{D-1}{z^2}-\frac{2(D-1)A'(z)}{z}+(D-1)A'(z)^2-\frac{\phi'(z)^2}{2(D-2)}  \right)\nonumber \\
+\frac{e^{-2A(z)}z^2f(z)A_{t}'(z)^2}{2(D-2)L^2} +\frac{e^{2A(z)}L^2V(z)}{(D-2)z^2} -(D-3)\kappa =0 \,.
\label{Einsteinzz}
\end{eqnarray}
\begin{eqnarray}
 x_{i}x_{i}: \ g''(z) + 2(D-2) g'(z) \left(A'(z)-\frac{1}{z}\right) -\frac{e^{-2A(z)}z^2f(z)A_{t}'(z)^2}{2L^2} +\frac{e^{2A(z)}L^2V(z)}{z^2} \nonumber \\
  + (D-2) g(z)\left(\frac{D-1}{z^2} -2(D-3)\frac{A'(z)}{z} +(D-3)A'(z)^2 + \frac{\phi'(z)^2}{2(D-2)} + 2 A''(z) \right) \nonumber \\
  -(D-3)(D-4)\kappa =0 \,.
\label{Einsteinxixi}
\end{eqnarray}
These complicated looking expressions, however, can be rearranged into the following simpler forms, which are then much easier to analyse:
\begin{eqnarray}
g''(z) + (D-2)g'(z) \left(A'(z)-\frac{1}{z}\right) - \frac{e^{-2A(z)}z^2f(z)A_{t}'(z)^2}{L^2} + 2(D-3)\kappa = 0\,.
\label{EOM11}
\end{eqnarray}
\begin{eqnarray}
A''(z) - A'(z) \left(A'(z)-\frac{2}{z}\right)+\frac{\phi'(z)^2}{2(D-2)} = 0 \,.
\label{EOM22}
\end{eqnarray}
\begin{eqnarray}
\frac{g''(z)}{4g(z)}+\frac{(D-2)}{2}A''(z) + (D-2)^2 A'(z)\left(-\frac{1}{z}+\frac{A'(z)}{2}+\frac{3}{4(D-2)}\frac{g'(z)}{g(z)}  \right) \nonumber \\
 -\frac{3(D-2)}{4}\frac{g'(z)}{zg(z)} +\frac{e^{2A(z)}L^2V(z)}{2z^2g(z)} +\frac{(D-1)(D-2)}{2z^2} -\frac{(D-3)^2 \kappa}{2g(z)}   = 0 \,.
\label{EOM33}
\end{eqnarray}
Similarly, we get the following equation of motion for the scalar field,
\begin{eqnarray}
 \phi ''(z) +\phi '(z) \left(\frac{g'(z)}{g(z)}+(D-2)A'(z)-\frac{D-2}{z}\right) + \frac{e^{-2A(z)}z^2 A_{t}'(z)^2}{2 L^2 g(z)}\frac{\partial f(\phi)}{\partial \phi}  \nonumber \\
    -\frac{L^2 e^{2A(z)}}{z^2 g(z)} \frac{\partial V(\phi)}{\partial \phi} =0\,.
\label{dilatonEOM}
\end{eqnarray}
and the equation of motion for the gauge field,
\begin{eqnarray}
A_{t}''(z)+ A_{t}'(z) \left(\frac{f'(z)}{f(z)}+(D-4)A'(z)-\frac{D-4}{z}\right) =0\,.
\label{MaxwellAtEOM}
\end{eqnarray}
Hence, we have a total of five equations of motion. However, it can be easily shown that only four of them are independent. Here, we choose the scalar equation (\ref{dilatonEOM}) as a constrained equation and consider the remaining equations as independent. Further, in order to solve these equations, we impose the following boundary conditions,
\begin{eqnarray}
&& g(0)=1 \ \ \text{and} \ \ g(z_h)=0, \nonumber \\
&& A_{t}(0)= \mu \ \ \text{and} \ \  A_{t}(z_h)=0, \nonumber \\
&& A(0) = 1, \nonumber \\
&& \phi(0)=0\,.
\label{boundaryconditions}
\end{eqnarray}
where $\mu$ is the chemical potential which, in the standard gauge/gravity duality language, is related to the near boundary expansion of the gauge field $A_t(z)$. Apart from these boundary conditions, we also demand that the scalar field $\phi$ remains real everywhere in the bulk.\\

We consider the following strategy to solve the Einstein-Maxwell-Scalar equations of motion simultaneously

\begin{itemize}
\item First, we solve eq.~(\ref{MaxwellAtEOM}) and obtain a solution for the gauge field $A_t(z)$ in terms of $A(z)$ and $f(z)$.

\item Using the obtained $A_t(z)$, we then solve eq.~(\ref{EOM11}) and find a solution for $g(z)$ in terms of $A(z)$ and $f(z)$.

\item Subsequently, we solve eq.~(\ref{EOM22}) and find $\phi'(z)$ in terms of $A(z)$.

\item Lastly, we solve eq.~(\ref{EOM33}) and obtain the scalar potential $V$ in terms of $A(z)$ and $g(z)$.
\end{itemize}

Adopting the above mentioned strategy and solving eq.~(\ref{MaxwellAtEOM}), we get the following solution for $A_t(z)$
\begin{eqnarray}
A_{t} (z) = C_{1} \int_0^z \, d\xi \frac{e^{-(D-4)A(\xi)} \xi^{(D-4)}}{f(\xi)} +C_2 \,.
\end{eqnarray}
where $C_1$ and $C_2$ are the integration constants and they are found to be
\begin{eqnarray}
C_2 = \mu,  \ \ \ \ \ C_1 = -\frac{\mu}{ \int_0^{z_h} \, d\xi \frac{e^{-(D-4)A(\xi)} \xi^{(D-4)}}{f(\xi)} }  \,.
\end{eqnarray}
by using the boundary conditions (eq.(\ref{boundaryconditions})). The solution for the gauge field $A_{t}(z)$ then becomes,
\begin{eqnarray}
A_{t} (z) = \mu \biggl[ 1 - \frac{\int_0^z \, d\xi \frac{e^{-(D-4)A(\xi)} \xi^{(D-4)}}{f(\xi)}}{\int_{0}^{z_h} \, d\xi \frac{e^{-(D-4)A(\xi)} \xi^{(D-4)}}{f(\xi)}} \biggr] = \tilde{\mu} \int_z^{z_h} \, d\xi \frac{e^{-(D-4)A(\xi)} \xi^{(D-4)}}{f(\xi)}  \,.
 \label{Atsol}
\end{eqnarray}
Now, using eq.~(\ref{Atsol}) into eq.~(\ref{EOM11}), we get the following solution for $g(z)$,
\begin{eqnarray}
& & g(z) =  C_4 + \int_0^z \, d\xi \ e^{-(D-2)A(\xi)} \xi^{(D-2)} \biggl[ C_{3} + \mathcal{K}(\xi) \biggr] \, \nonumber\\
& & \mathcal{K}(\xi)= \int \, d\xi \ \biggl[ \frac{\tilde{\mu}^2 \xi^{(D-4)} e^{-(D-4)A(\xi)}}{L^2 f(\xi)} -2(D-3)\kappa \frac{e^{(D-2)A(\xi)}}{\xi^{D-2}} \biggr]
\label{gsol}
\end{eqnarray}
where the constants $C_3$ and $C_4$ can again be fixed from eq.~(\ref{boundaryconditions}) and we get,
\begin{eqnarray}
C_4 = 1,  \ \ \ \ \ C_3 =- \frac{1+ \int_0^{z_h} \, d\xi e^{-(D-2)A(\xi)} \xi^{(D-2)} \mathcal{K}(\xi) }{ \int_0^{z_h} \, d\xi e^{-(D-2)A(\xi)} \xi^{(D-2)} }  \,.
\end{eqnarray}
Similarly, the scalar field $\phi$ can be solved in terms of $A(z)$ from eq.~(\ref{EOM22})
\begin{eqnarray}
\phi(z) = \int dz \sqrt{2(D-2)\biggl[-A''(z) + A'(z) \left(A'(z)-\frac{2}{z}\right) \biggr]} + C_{5}
\label{phisol}
\end{eqnarray}
where the constant $C_{5}$ will be fixed demanding that $\phi$ vanishes near the asymptotic boundary, \textit{i.e.} $\phi |_{z=0}\rightarrow 0$. And finally, we can find the potential $V$ from eq. (\ref{EOM33}),
\begin{eqnarray}
V(z) = -\frac{2z^2g(z)e^{-2A(z)}}{L^2} \biggl[\frac{(D-1)(D-2)}{2z^2}-\frac{3(D-2)}{4}\frac{g'(z)}{zg(z)}+\frac{g''(z)}{4g(z)}+\frac{(D-2)}{2}A''(z) \nonumber\\
 + (D-2)^2 \left(-\frac{1}{z}+\frac{A'(z)}{2}+\frac{3}{4(D-2)}\frac{g'(z)}{g(z)}  \right)A'(z) -\frac{(D-3)^2\kappa}{2g(z)}   \biggr]   \,.
\label{Vsol}
\end{eqnarray}
It is clear from the above equations that the Einstein-Maxwell-Scalar gravity system of eq.~(\ref{actionEF}) can be solved analytically in terms of two arbitrary functions, \textit{i.e.} scale function $A(z)$ and coupling $f(z)$. Correspondingly, depending upon the forms of $A(z)$ and $f(z)$, a charged hairy black hole solution with various horizon topologies can be constructed analytically in all spacetime dimensions. However, different forms of $A(z)$ and $f(z)$ will correspond to different physically allowed charged hairy black hole solutions, as different forms of these functions will correspond to different potentials $V(z)$. We, therefore, have an infinite family of analytic charged hairy black hole solutions for the Einstein-Maxwell-Scalar gravity system of eq. (2.1).

Nonetheless, in the context of gauge/gravity duality, the forms of $A(z)$ and $f(z)$ are generally fixed by taking inputs from the dual boundary theory. For instance, in the field of holographic QCD model building, the forms of $A(z)$ and $f(z)$ are generally fixed by requiring the dual boundary theory to exhibit physical QCD properties such as confinement/deconfinement phase transition, linear Regge trajectory for the excited meson mass spectrum, confinement in the quark sector etc. In recent years, a particularly interesting form of $A(z) = -a z^n$, with $n$ being non-negative, has been extensively used in the literature, as the corresponding boundary theory imitates many of lattice QCD properties \cite{Dudal:2017max,Dudal:2018ztm,Gursoy:2007er,Gursoy:2008za
}. Moreover, one can also put a constraint on the value of parameter $a$, such as $a>0$, by demanding the thermal-AdS/black hole phase transition temperature in gravity side (or the dual confinement/deconfinement phase transition temperature in the boundary side) to be equal to lattice QCD predicted value, and on $n$, such as $n>1$, by demanding confinement behaviour at low temperatures \cite{Dudal:2017max,Dudal:2018ztm}. In this paper, however, we will take a more liberal approach and consider various values of $a$ and $n$ (hence different scalar potential) to investigate the effects of scalar hair on the black hole physics in different dimensions.

Another reason for choosing $A(z)=-az^n$ is that it ensures that the constructed spacetime geometry asymptotes to AdS at the boundary $z\rightarrow 0$. Note that near the
asymptotic boundary, we have
\begin{eqnarray}
& & V(z)|_{z\rightarrow 0} = -\frac{D(D-1)}{L^2} + \frac{m^2\phi^2}{2}+\dots \nonumber\\
 & &  V(z)|_{z\rightarrow 0} =  2\Lambda + \frac{M^2\phi^2}{2}+\dots   \,.
\label{Vsolexp}
\end{eqnarray}
where $\Lambda=-\frac{D(D-1)}{L^2}$ is as usual the negative cosmological constant. This, along with the fact that $g(z)|_{z\rightarrow 0}=1$, indeed ensures that the geometry asymptotes to AdS at the boundary. $M^2$ is the mass of the scalar field, and it satisfies the Breitenlohner-Freedman bound for stability in AdS space \textit{i.e} $M^2\geq-(D-1)^2/4$ \cite{Breitenlohner:1982jf}. Moreover, from the gauge/gravity duality point of view, the Gubser criterion in order to have a well defined dual boundary theory is also respected by the gravity system under consideration. In particular, $V(0)\geq V(z)$ \textit{i.e.} the potential is bounded from above by its UV boundary value \cite{Gubser:2000nd}.

Similarly, one has the freedom to fix the coupling function $f(z)$ by taking inputs from the dual boundary theory as well. For example, $f(z)$ can be constrained by demanding the meson mass spectrum of the dual boundary theory to lie on the linear Regge trajectory. In five dimensions, a simpler form like $f(z)=e^{-c z^2 - A(z)}$ can do this job. However, in order to have a better control over the integrals that appear in the expressions of $g(z)$, $\phi(z)$, $A_t(z)$ and $V(z)$, and without worrying too much about the dual boundary theory, here we consider a particular simpler form $f(z)=e^{-(D-4)A(z)}$. Again, this form does not modify the asymptotic structure of the spacetime.\\

It is important to emphasize again that eqs.~(\ref{Atsol})-(\ref{Vsol}) are a solution to the Einstein-Maxwell-Scalar action (\ref{actionEF}) for any $A(z)$ and $f(z)$. We, therefore, have an infinite family of analytic charged hairy black hole solutions. These different black hole solutions however correspond to different scalar potentials $V$ (and
therefore to different Einstein-Maxwell-Scalar actions) as different forms of these functions will give different $V(z)$. Nonetheless, once we have fixed the forms of $A(z)$ and $f(z)$ then the form of $V(z)$ is also fixed, and in return, eqs.~(\ref{Atsol})-(\ref{Vsol}) provide a self-consistent solution to a particular Einstein-Maxwell-Scalar action with
predetermined $A(z)$, $f(z)$ and $V(z)$.\\

Before ending this section, we would like to mention that there exists another allowed solution to the Einstein-Maxwell-Scalar equations of motion, \textit{i.e.} one without the horizon corresponding to thermal-AdS. The thermal-AdS solution can be obtained by taking the limit $z_h\rightarrow \infty$ in the black hole solution given above,\textit{ i.e.} $g(z)=1$. Although, this thermal-AdS solution again goes to AdS asymptotically, however depending on the form of $A(z)$, it can have a non-trivial structure in the bulk spacetime. As we will see shortly, depending on the magnitude of parameters $\kappa$, $a$ and $n$, there can be a Hawking-Page type thermal-AdS/black hole phase transition between these two solutions.\\

Let us also record the expressions of black hole entropy\footnote{It is actually the black hole entropy per unit transverse area.} and temperature, which will be useful in the discussion of the black hole thermodynamics later on
\begin{eqnarray}
& & S_{BH}=\frac{L^{D-2}e^{(D-2)A(z_h)}}{4 G_D z_{h}^{D-2}} \, , \nonumber\\
& & T= \frac{z_{h}^{D-2} e^{-(D-2)A(z_h)}}{ 4 \pi } \biggl[-\mathcal{K}(z_h) +  \frac{1+ \int_0^{z_h} \, d\xi e^{-(D-2)A(\xi)} \xi^{(D-2)} \mathcal{K}(\xi) }{ \int_0^{z_h} \, d\xi e^{-(D-2)A(\xi)} \xi^{(D-2)} }  \biggr] \,.
\label{STexp}
\end{eqnarray}

\section{Stability and black hole thermodynamics}
In this section, we discuss the thermodynamics of the charged hairy gravity system constructed in the previous section. There are various parameters, such as $a$, $n$, $\kappa$, $\mu$ and $D$ in our model, and all these parameters greatly modify the thermodynamic of the black hole. In particular, important physical properties such as hairy/RN-AdS black hole phase transition, critical point etc are all sensitive to these parameters.

\subsection{Planar horizon: $\kappa=0$}

\subsubsection{Case: n=1}
\begin{figure}[ht]
	\subfigure[]{
		\includegraphics[scale=0.4]{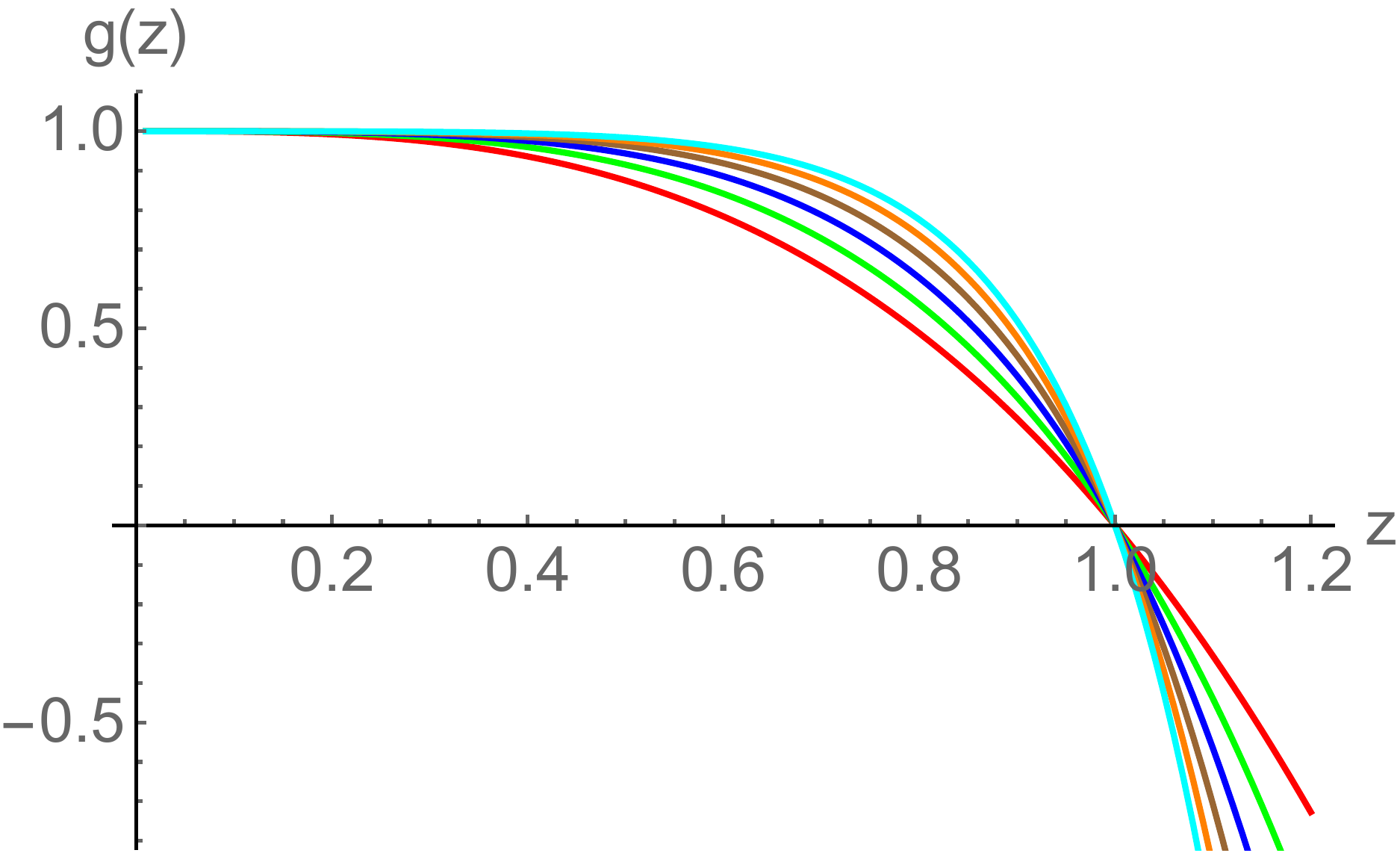}
	}
	\subfigure[]{
		\includegraphics[scale=0.4]{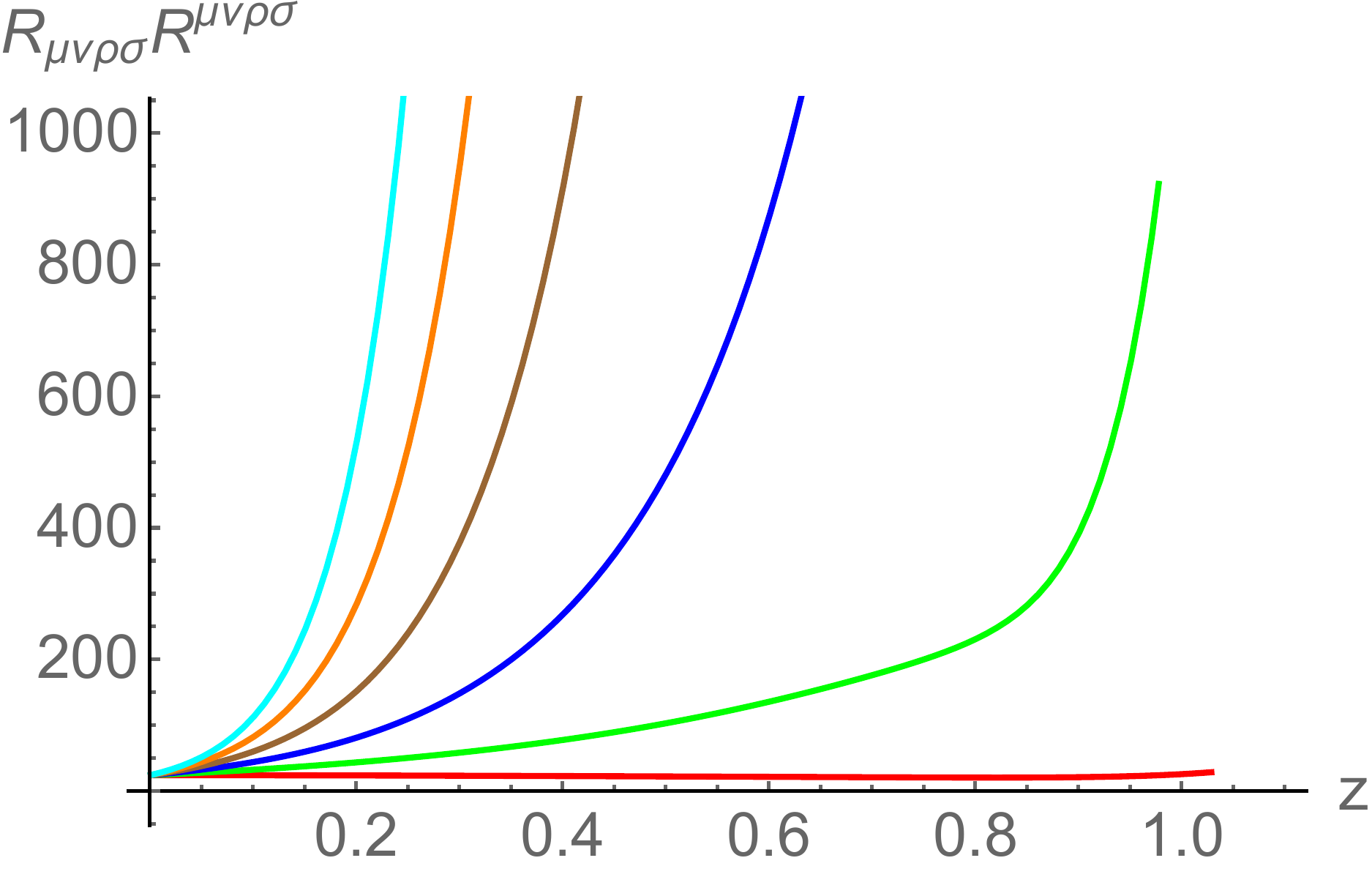}
	}
	\subfigure[]{
		\includegraphics[scale=0.4]{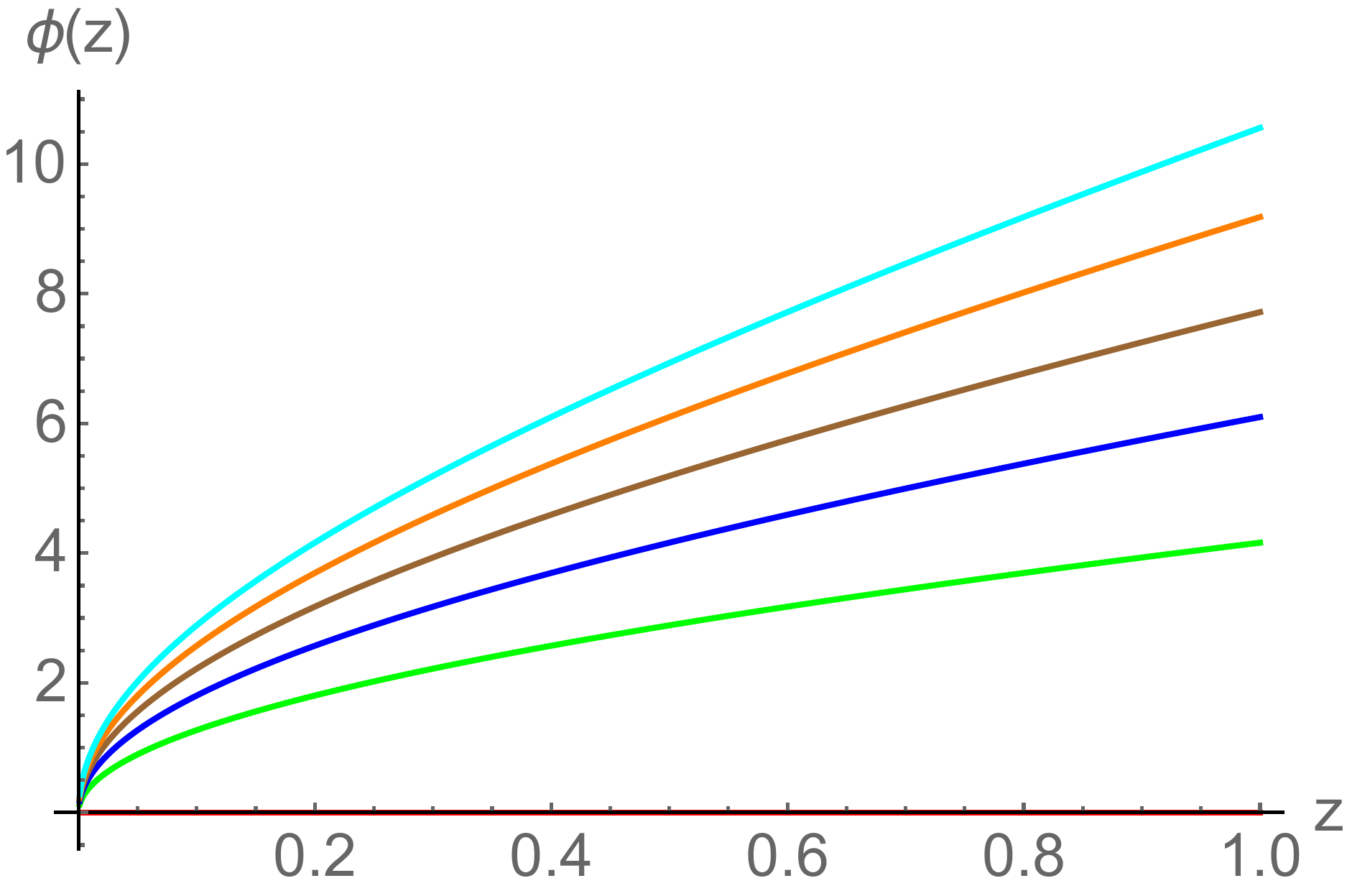}
	}
\subfigure[]{
		\includegraphics[scale=0.4]{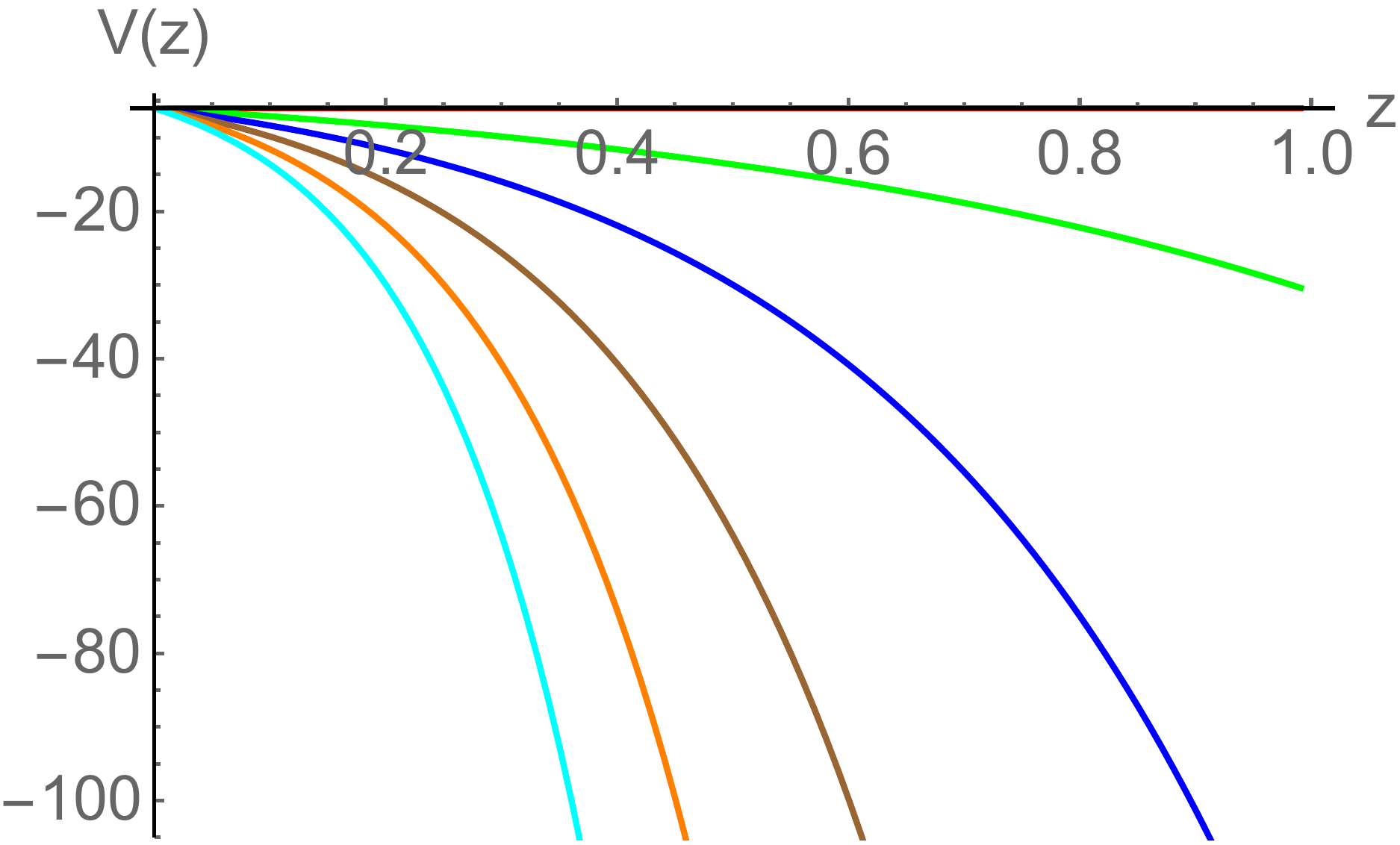}
	}
	\caption{\small The behavior of $g(z)$, $R_{\mu\nu\rho\lambda}R^{\mu\nu\rho\lambda}$, $\phi(z)$ and $V(z)$ for different values of $a$. Here $n=1$, $z_h=1$, $\mu=0.1$, $\kappa=0$ and $D=4$ are used. Red, green, blue, brown, orange and cyan curves correspond to $a=0$, $0.5$, $1.0$, $1.5$, $2.0$ and $2.5$ respectively.}
	\label{zvsgvsamuPt1zh1D4n1}
\end{figure}

Let us first discuss the thermodynamics of the gravity system with $n=1$, corresponding to $A(z)=-az$, for the planar horizon. With $A(z)=-az$, the expressions for scalar field $\phi(z)$  and  metric function $g(z)$ in four dimensions reduce to,
\begin{eqnarray}
\resizebox{0.4\hsize}{!}{$
\phi(z)= 2 \sqrt{a z (a z+2)}+4 \sinh ^{-1}\left(\frac{\sqrt{a z}}{\sqrt{2}}\right) \,. $}
\label{phiD4n1planar}
\end{eqnarray}
\begin{eqnarray}
\resizebox{1.0\hsize}{!}{$
g(z)= \frac{e^{2az}}{2} \biggl[\frac{\left(-4 a^2 z^2+4 a z+2 e^{-2 a z}-2\right) \left(\frac{\mu ^2 \left(\left(4 a^3 z_h^3-6
   a^2 z_h^2+6 a z_h-3\right) \left(\sinh \left(2 a z_h\right)+\cosh \left(2 a
   z_h\right)\right)+3\right)}{8 a^4 z_h^2}+1\right)}{\left(2 a^2 z_h^2-2 a z_h+1\right) \left(\sinh
   \left(2 a z_h\right)+\cosh \left(2 a z_h\right)\right)-1}\biggr] $}  \nonumber\\
  \resizebox{0.5\hsize}{!}{$ +1 + \frac{e^{2az}}{16 a^4}\biggl[\frac{\mu ^2 \left(8 a^3 z^3-12 a^2 z^2+12 a z+6 e^{-2 a z}-6\right)}{z_{h}^2} \biggr]
   \,. $}
\label{gD4n1planar}
\end{eqnarray}
Similarly, an analytical expression for the potential $V(z)$ can be found.  However, the expression is too long and not very illuminating, therefore we skip to reproduce it here for brevity. In Figure~\ref{zvsgvsamuPt1zh1D4n1}, we have plotted these functions for different values of $a$ in four spacetime dimensions. Here, we have shown results for a particular value of the chemical potential $\mu=0.1$ and horizon radius $z_h=1$, however, analogous results hold for other values of $\mu$ and $z_h$ as well. The metric function $g(z)$ changes sign at $z=z_h$ for all values of $a$, signalling the presence of a horizon. Additionally, we have analysed the behaviour of the Kretschmann scalar $R_{\mu\nu\rho\lambda}R^{\mu\nu\rho\lambda}$. As shown in Figure~\ref{zvsgvsamuPt1zh1D4n1}, it is finite everywhere outside the horizon, indicating no curvature singularity in the bulk spacetime.

\begin{figure}[h!]
\begin{minipage}[b]{0.5\linewidth}
\centering
\includegraphics[width=2.8in,height=2.3in]{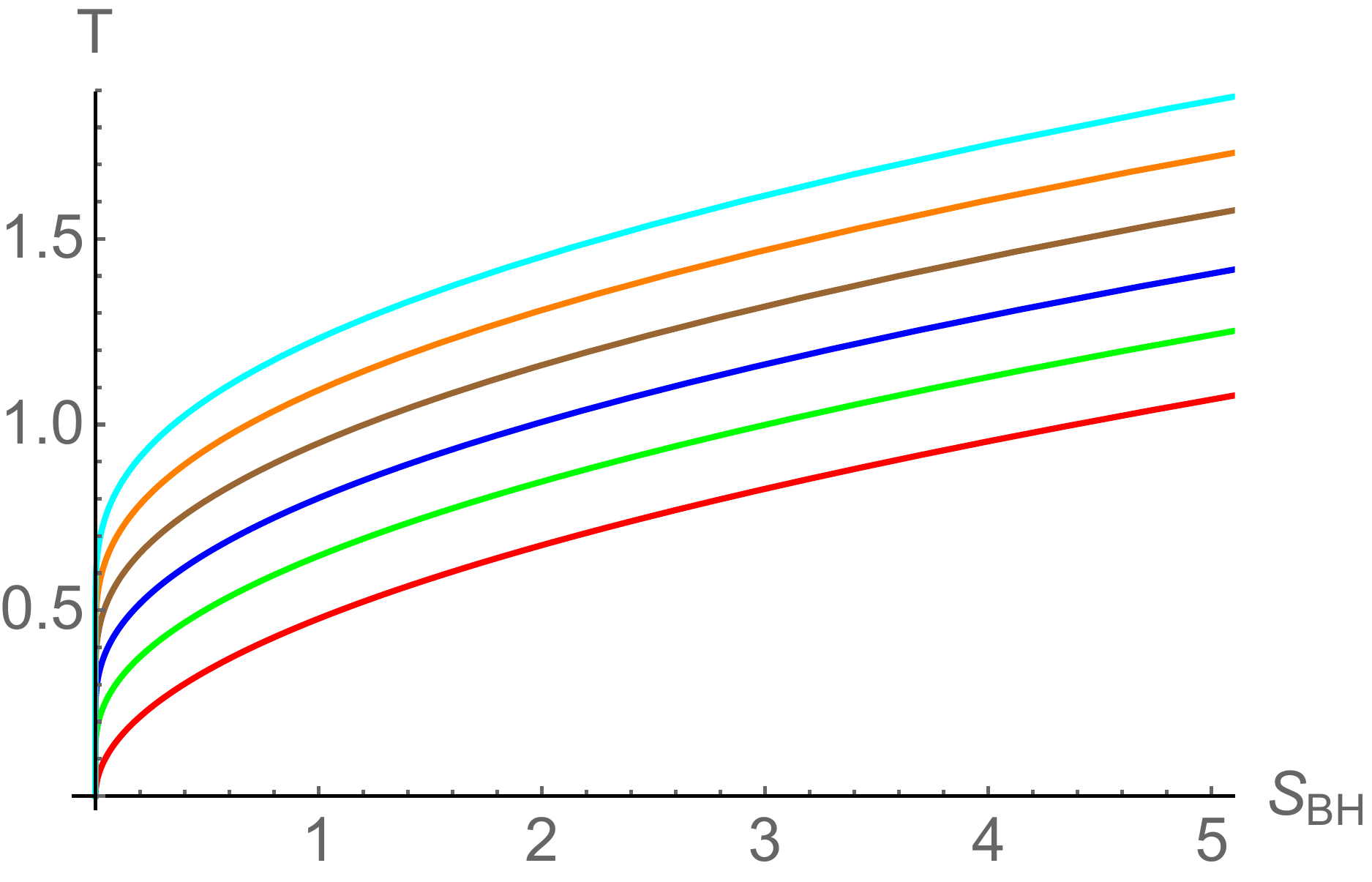}
\caption{ \small Hawking temperature $T$ as a function of black hole entropy $S_{BH}$ for various values $a$.  Here $n=1$, $\mu=0.1$, $\kappa=0$ and $D=4$ are used. Red, green, blue, brown, orange and cyan curves correspond to $a=0$, $0.5$, $1.0$, $1.5$, $2.0$ and $2.5$ respectively. }
\label{SBHvsTvsamuPt1D4n1}
\end{minipage}
\hspace{0.4cm}
\begin{minipage}[b]{0.5\linewidth}
\centering
\includegraphics[width=2.8in,height=2.3in]{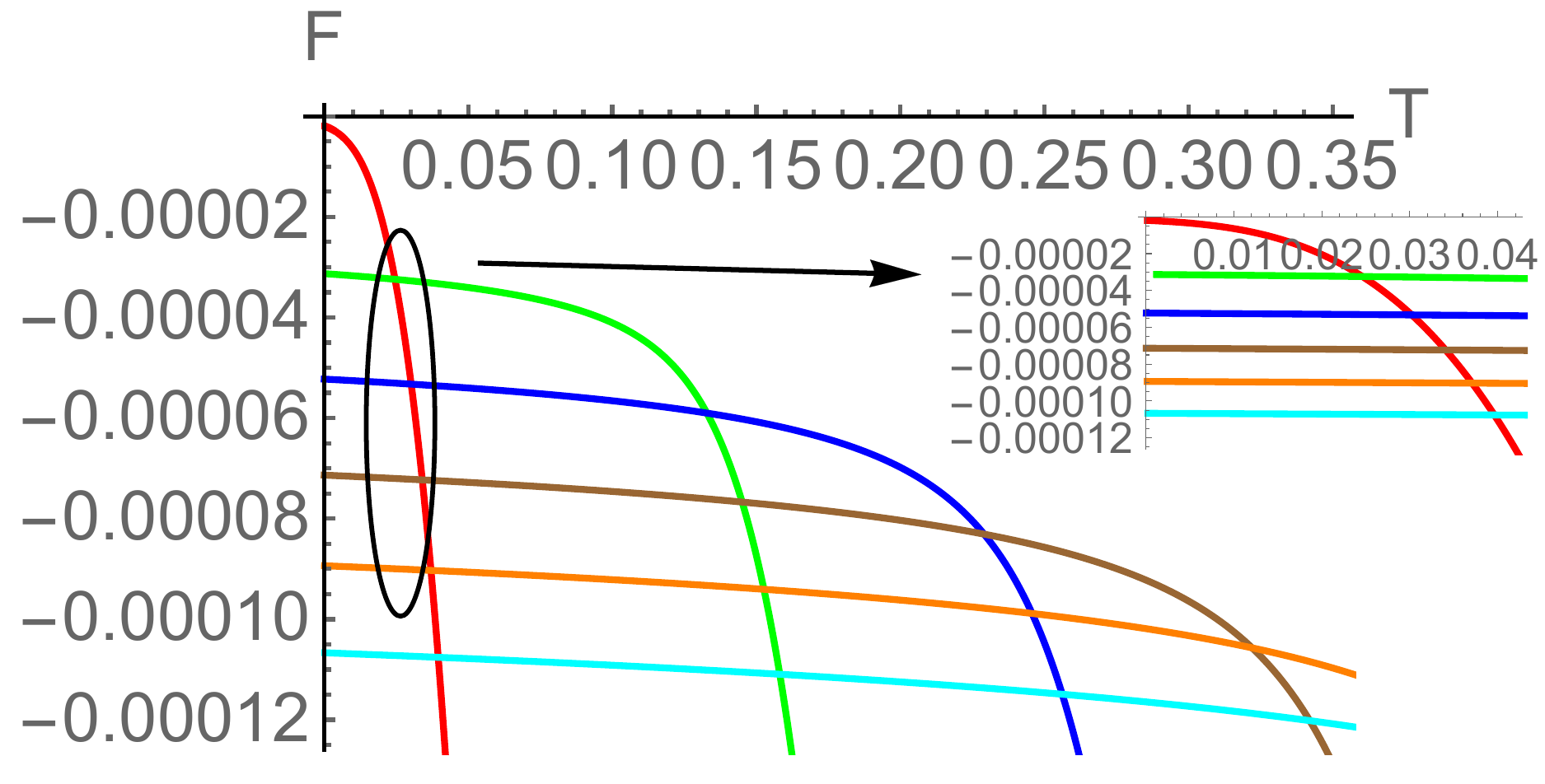}
\caption{\small Normalised Gibbs free energy $F$ as a function of Hawking temperature $T$ for various values $a$.  Here $n=1$, $\mu=0.1$, $\kappa=0$ and $D=4$ are used. Red, green, blue, brown, orange and cyan curves correspond to $a=0$, $0.5$, $1.0$, $1.5$, $2.0$ and $2.5$ respectively.}
\label{TvsFvsamuPt1D4n1}
\end{minipage}
\end{figure}

Further, we have analysed the scalar field behaviour and found it to be real and regular everywhere outside the horizon. As can be seen from eq.~(\ref{phiD4n1planar}), the scalar field goes to zero only at the asymptotic AdS boundary.  The finiteness of scalar field at and outside the horizon suggests the existence of a well behaved planar charged hairy black hole solution in the asymptotic AdS space in our model. The stability of the charged hairy black hole phase against the non-hairy RN-AdS black hole phase will be discussed shortly when we will analyse its free energy. Notice also that the non-zero values of the scalar field appear only when $a\neq0$, and it vanishes when $a=0$. This is consistent with our expectation that as $a\rightarrow 0$ one should get back the standard RN-AdS solution. This can be easily checked by taking the limit $a\rightarrow0$ in $g(z)$, in which case it simply reduces to planar RN-AdS expression. Similarly, the potential is regular everywhere and asymptotes to $V(0)=-6/L^2$ at the boundary for all values of $a$. $V(z)$ is constant for $a=0$, whereas for finite $a$, it decreases as $z$ increases. Importantly, the potential is bounded from above by its UV boundary value for all $a$.

In higher dimensions, analogous expressions for $g(z)$, $\phi(z)$ and $V(z)$ can again be obtained analytically.  The above-mentioned results for $g(z)$, $R_{\mu\nu\rho\lambda}R^{\mu\nu\rho\lambda}$, $\phi(z)$ and $V(z)$ persist in higher dimensions as well. In particular, there is no spacetime singularity as the Kretschmann scalar is always finite. Similarly, the scalar field is always real and regular outside the horizon. The potential again is bounded from above by its UV boundary value for all $a$ and asymptotes to a constant value \textit{i.e.}, $V(z)|_{z\rightarrow 0} =  2\Lambda$, in all dimensions.

In order to analyse the thermodynamic stability of the charged hairy black holes, it is important to investigate the local stability of these black holes. The local stability is measured by the response of the equilibrium system under a small fluctuation in thermodynamical variables, and in the grand canonical ensemble, it is quantified by the positivity of the specific heat at constant potential $C=T(\partial S_{BH}/\partial T)$, \textit{i.e.} the condition $C>0$ ensures the local stability of the thermodynamic system.  In Figure~\ref{SBHvsTvsamuPt1D4n1}, the variation of Hawking temperature with respect to black hole entropy for different values of $a$ is shown. We see that the slope in the $S_{BH}-T$ plane is always positive, indicating that the constructed hairy black holes are thermodynamically stable as the specific heat is always positive.

\begin{figure}[h!]
\centering
\includegraphics[width=2.8in,height=2.3in]{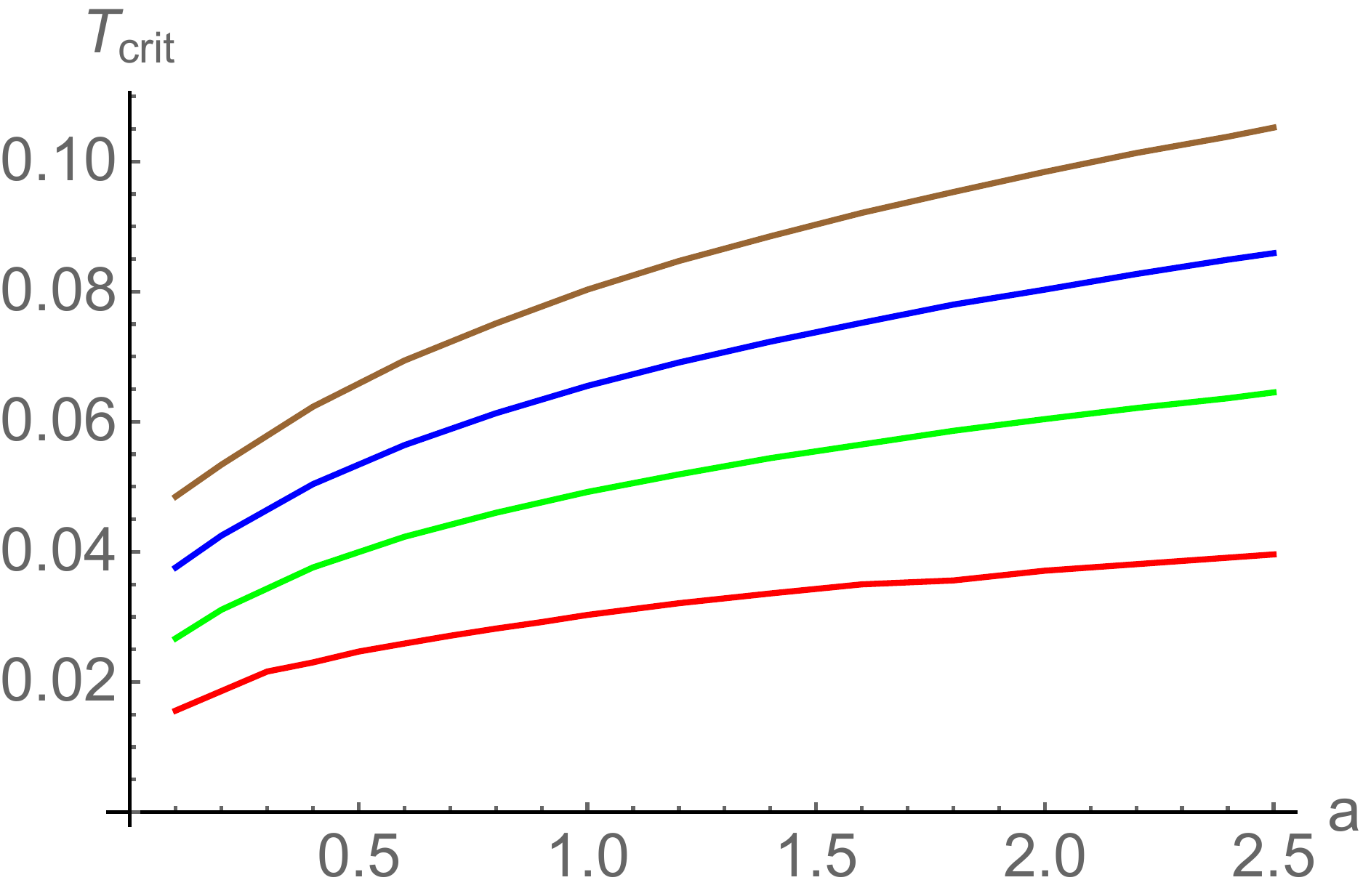}
\caption{\small The hairy/non-hairy black hole critical temperature $T_{crit}$ as a function $a$ for various values $\mu$. Here $n=1$, $\kappa=0$ and $D=4$ are used. Red, green, blue and brown curves correspond to $\mu=0.1$, $0.2$, $0.3$ and $0.4$ respectively.}
\label{avsTcritvsmuD4n1}
\end{figure}

To further investigate the thermodynamic stability of the charged hairy black hole, we need to study its free energy behaviour. In Figure~\ref{TvsFvsamuPt1D4n1}, the Gibbs free energy $F$ as a function of hawking temperature $T$ is shown\footnote{Here and in subsequent figures, $F$ is measured in units of the transverse area.}. Here, we have normalised the free energy with respect to the thermal AdS. We find that the free energy of planar RN-AdS black hole (red line, corresponding to $a=0$) is smaller than the planar charged hairy black hole at higher temperatures whereas the free energy of planar charged hairy black hole is smaller than the planar RN-AdS black hole at lower temperatures. This suggests that although the planar RN-AdS black hole is more stable at higher temperatures, however, it is the planar hairy black hole configuration which is thermodynamically more favourable at low temperatures. The critical temperature $T_{crit}$ at which a transition from planar RN-AdS to charged hairy black hole takes place increases with $a$, implying that the temperature range for which hairy black hole configuration remains stable increases with $a$. Additionally, we also find that $T_{crit}$ increases with $\mu$. This suggests that the stability of the hairy black hole phase enhances at higher temperatures as $\mu$ increases. The complete dependence of $T_{crit}$ on $a$ and $\mu$ is shown in Figure~\ref{avsTcritvsmuD4n1}.  The free energy analysis, therefore, does imply the existence of a well behaved and thermodynamically stable charged hairy black hole solution in asymptotically AdS spaces in our model.

\begin{figure}[h!]
\begin{minipage}[b]{0.5\linewidth}
\centering
\includegraphics[width=2.8in,height=2.3in]{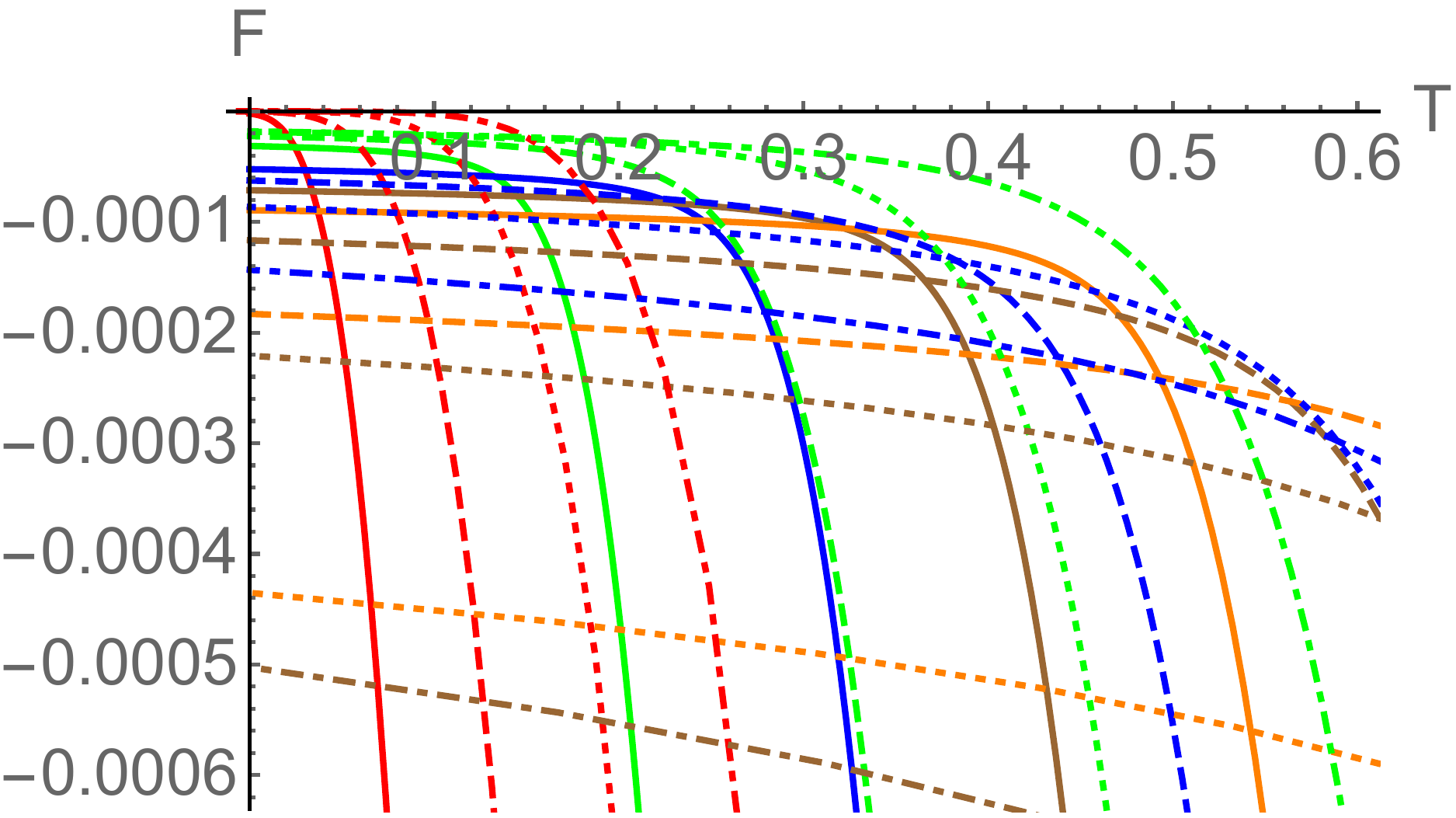}
\caption{ \small  Normalised Gibbs free energy $F$ as a function of Hawking temperature $T$ for various values of $a$ and $D$.  Here $n=1$, $\mu=0.1$ and $\kappa=0$ are used. Red, green, blue, brown and orange curves correspond to $a=0$, $0.5$, $1.0$, $1.5$ and $2.0$ respectively. Solid, dashed, dotted, dot-dashed curves correspond to $D=4$, $5$, $6$ and $7$ respectively. }
\label{TvsFvsavsDmuPt1n1}
\end{minipage}
\hspace{0.4cm}
\begin{minipage}[b]{0.5\linewidth}
\centering
\includegraphics[width=2.8in,height=2.3in]{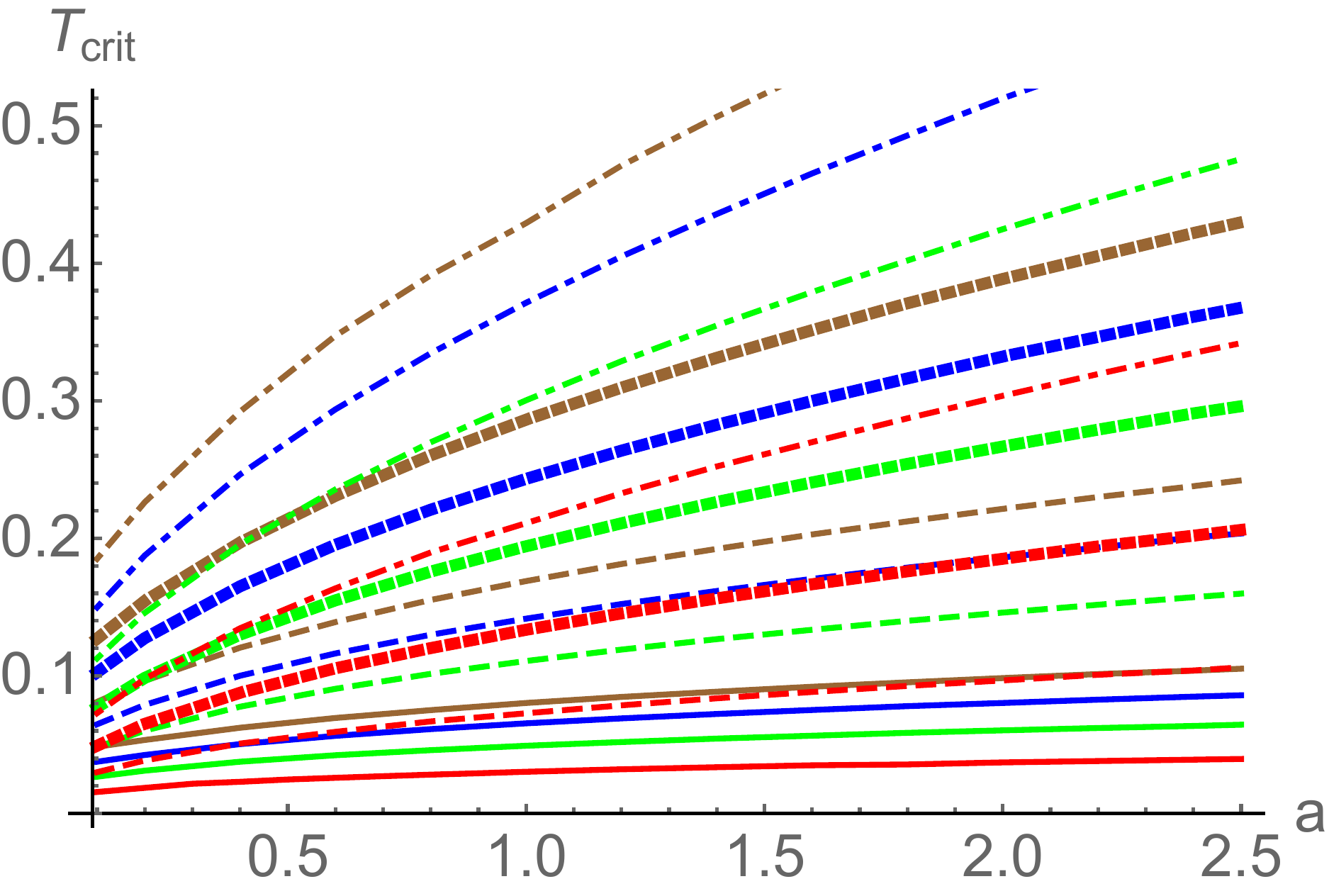}
\caption{\small  The critical temperature $T_{crit}$ as a function $a$ for various values of $\mu$ and $D$. Here $n=1$ and $\kappa=0$ are used. Red, green, blue and brown curves correspond to $\mu=0.1$, $0.2$, $0.3$ and $0.4$ respectively. Solid, dashed, dotted, and dot-dashed lines correspond to $D=4$, $5$, $6$ and $7$ respectively. }
\label{avsTcritvsmuvsDn1}
\end{minipage}
\end{figure}

The above-mentioned results for the existence of stable charged hairy planar black hole solution remain true in higher dimensions as well. In particular, there again exists a critical temperature below which the charged hairy planar black hole phase becomes more favourable than the planar RN-AdS phase.  Similarly, the critical temperature depends non-trivially on $a$ and $\mu$. The difference arises in the magnitude of $T_{crit}$, which increases as $D$ increases for the same value of $a$ and $\mu$. The possibility of scalar hair stabilization outside the horizon, therefore, increases in higher dimensions. The temperature dependence of free energy and the complete phase diagram displaying the dependence of $T_{crit}$ on $\mu$, $a$ and $D$ are shown in Figures~\ref{TvsFvsavsDmuPt1n1} and \ref{avsTcritvsmuvsDn1}.

It is important to emphasize that the planar hairy black hole becomes favourable only when $\mu\neq0$. For $\mu=0$, the free energy of planar RN-AdS black hole is always smaller than the planar hairy black hole. Therefore, the uncharged hairy black hole is always thermodynamically unfavourable in comparison with the Schwarzschild AdS black hole.  Moreover, note that $F$ is always negative and goes to zero only when $z_h\rightarrow \infty$. This indicates that black hole phases, both RN-AdS and hairy, are thermodynamically more favourable than the thermal-AdS phase.

\subsubsection{Case: n=2}
\begin{figure}[ht]
	\subfigure[]{
		\includegraphics[scale=0.4]{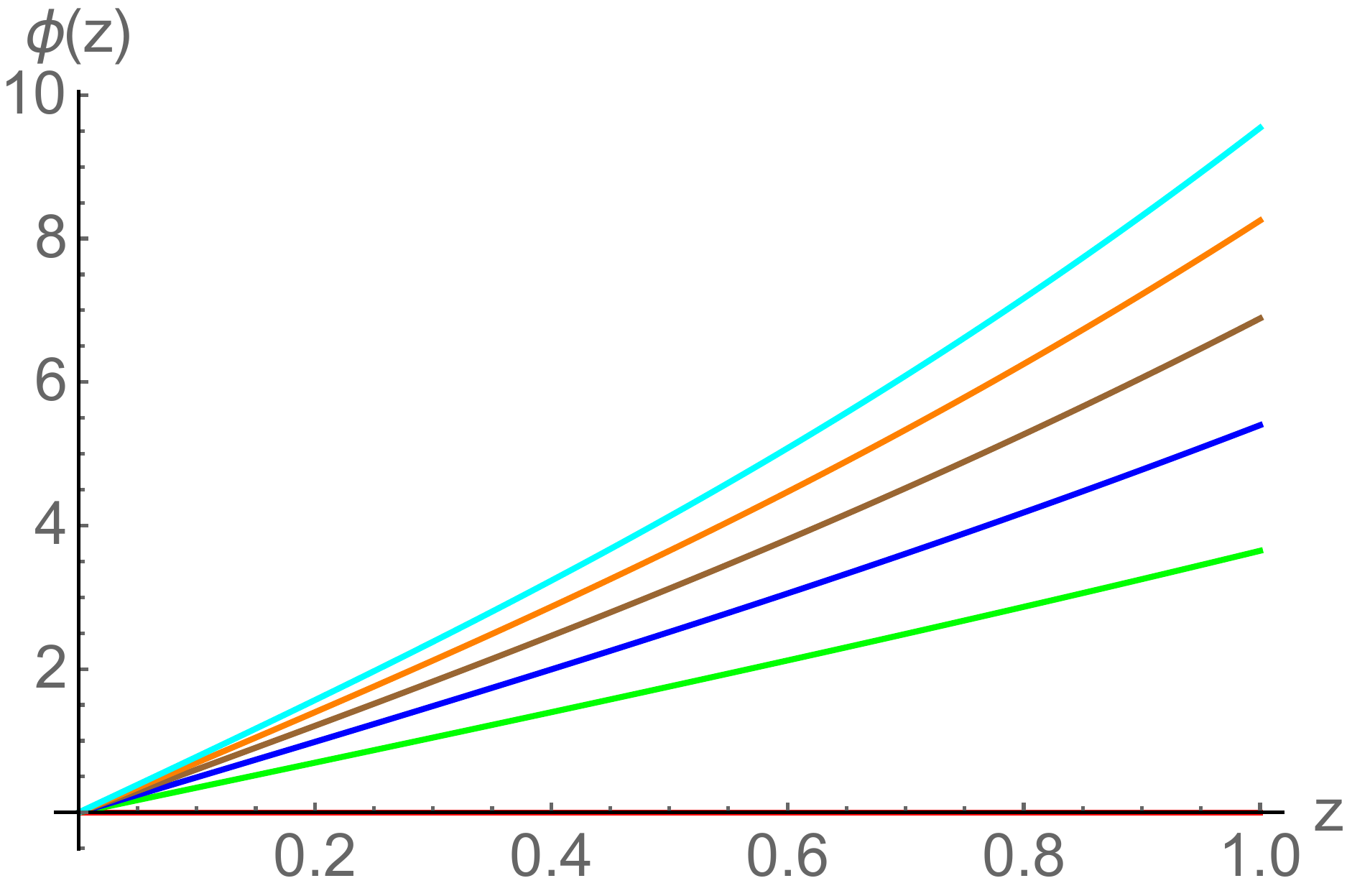}
	}
\subfigure[]{
		\includegraphics[scale=0.4]{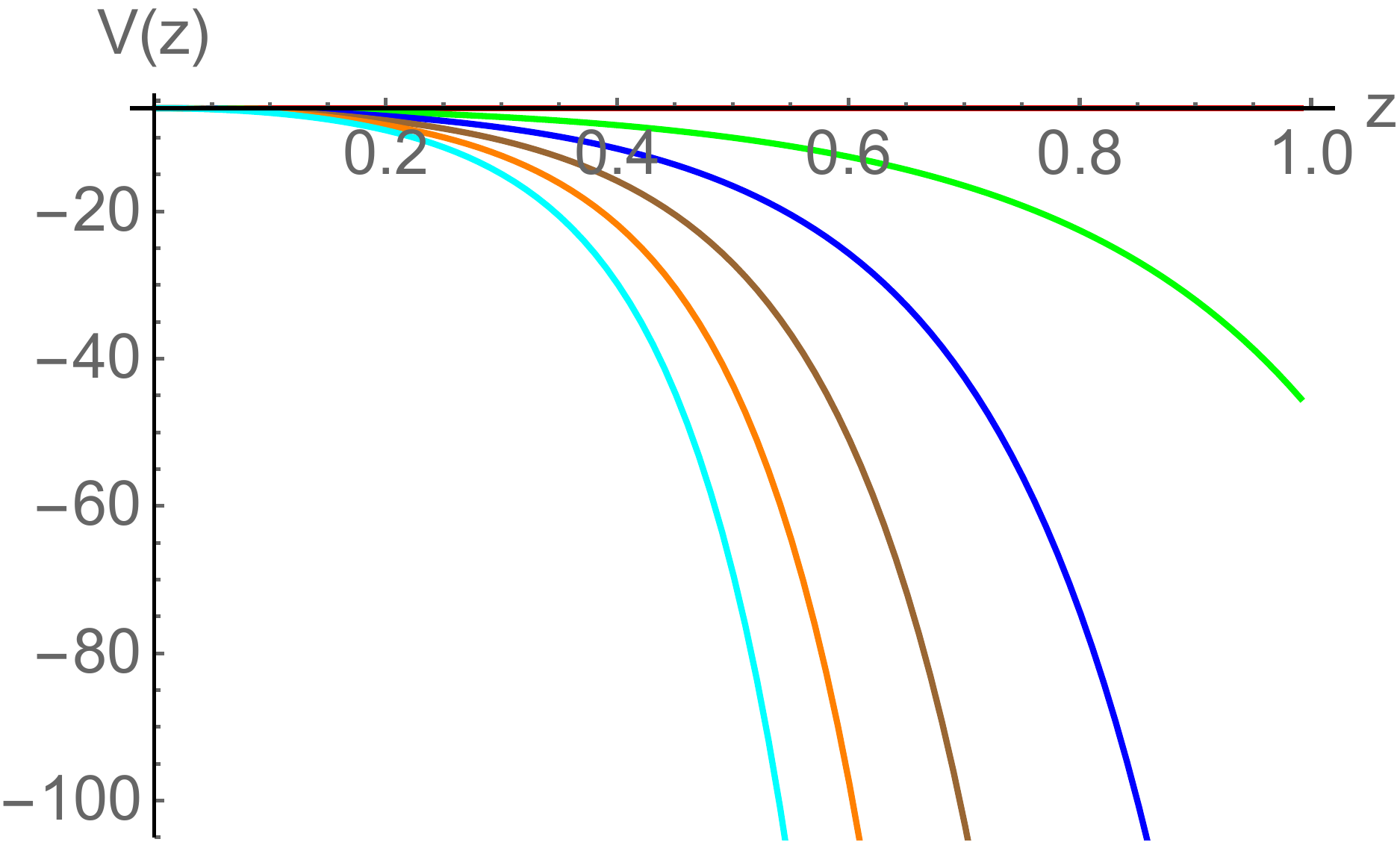}
	}
	\caption{\small The behavior of $\phi(z)$ and $V(z)$ for different values of $a$. Here $n=2$, $z_h=1$, $\mu=0.1$, $\kappa=0$ and $D=4$ are used. Red, green, blue, brown, orange and cyan curves correspond to $a=0$, $0.5$, $1.0$, $1.5$, $2.0$ and $2.5$ respectively.}
	\label{zvsgvsamuPt1zh1D4n2}
\end{figure}

Let us now discuss the stability and thermodynamics of the gravity system with $n=2$, corresponding to $A(z)=-az^2$. Similar to $n=1$ case, analytic expressions for $g(z)$, $\phi(z)$ and $V(z)$ in different dimensions can be obtained analytically for $n=2$ case as well. The expressions for $\phi(z)$ and $g(z)$ in four dimensions are given by,
\begin{eqnarray}
\phi(z)=  z \sqrt{2 a \left(2 a z^2+3\right)}+3 \log \left(\frac{ \sqrt{2a \left(2 a z^2+3\right)}+2
   a z}{\sqrt{6 a} }\right) \,.
\label{phiD4n2planar}
\end{eqnarray}
\begin{eqnarray}
g(z)=   1 + \frac{\mu ^2 \left[e^{2 a z^2} \left(2 a z^2-1\right)+1\right]}{8 a^2 z_h^2} + \biggl[ \left(4 \sqrt{a} z e^{2 a z^2}-\sqrt{2 \pi } \ \text{Erfi}\left(\sqrt{2a} z \right)\right)
   \left(\sinh \left(a z_h^2\right)-\cosh \left(a z_h^2\right)\right) \nonumber \\
   \times \frac{a \left(4 a+\mu ^2\right) z_h^2 \cosh \left(a z_h^2\right)-\left(4 a^2 z_h^2-a \mu ^2 z_h^2+\mu
   ^2\right) \sinh \left(a z_h^2\right)}{8 a^2 z_h^2 \left(2 \sqrt{a} z_h-\sqrt{2}
  \  \text{DawsonF}\left(\sqrt{2 a} z_h\right)\right)} \biggr] \,.
\label{gD4n1p2anar}
\end{eqnarray}
where Erfi(x) and DawsonF(x) are the imaginary error function and Dawson integral respectively\footnote{The Dawson integral is define by DawsonF$(x)=e^{-x^2}\int_{0}^{x} dy \ e^{y^2}$.}. The results for $\phi(z)$ and $V(z)$ are shown in Figure~\ref{zvsgvsamuPt1zh1D4n2}. The scalar field is again regular near the horizon and goes to zero only at the asymptotic boundary. Correspondingly, the potential asymptotes to a constant value $V(z)|_{z\rightarrow 0} = 2\Lambda$ near the boundary and is bounded from above. Similarly, no spacetime singularity exists outside the horizon.

\begin{figure}[h!]
\begin{minipage}[b]{0.5\linewidth}
\centering
\includegraphics[width=2.8in,height=2.3in]{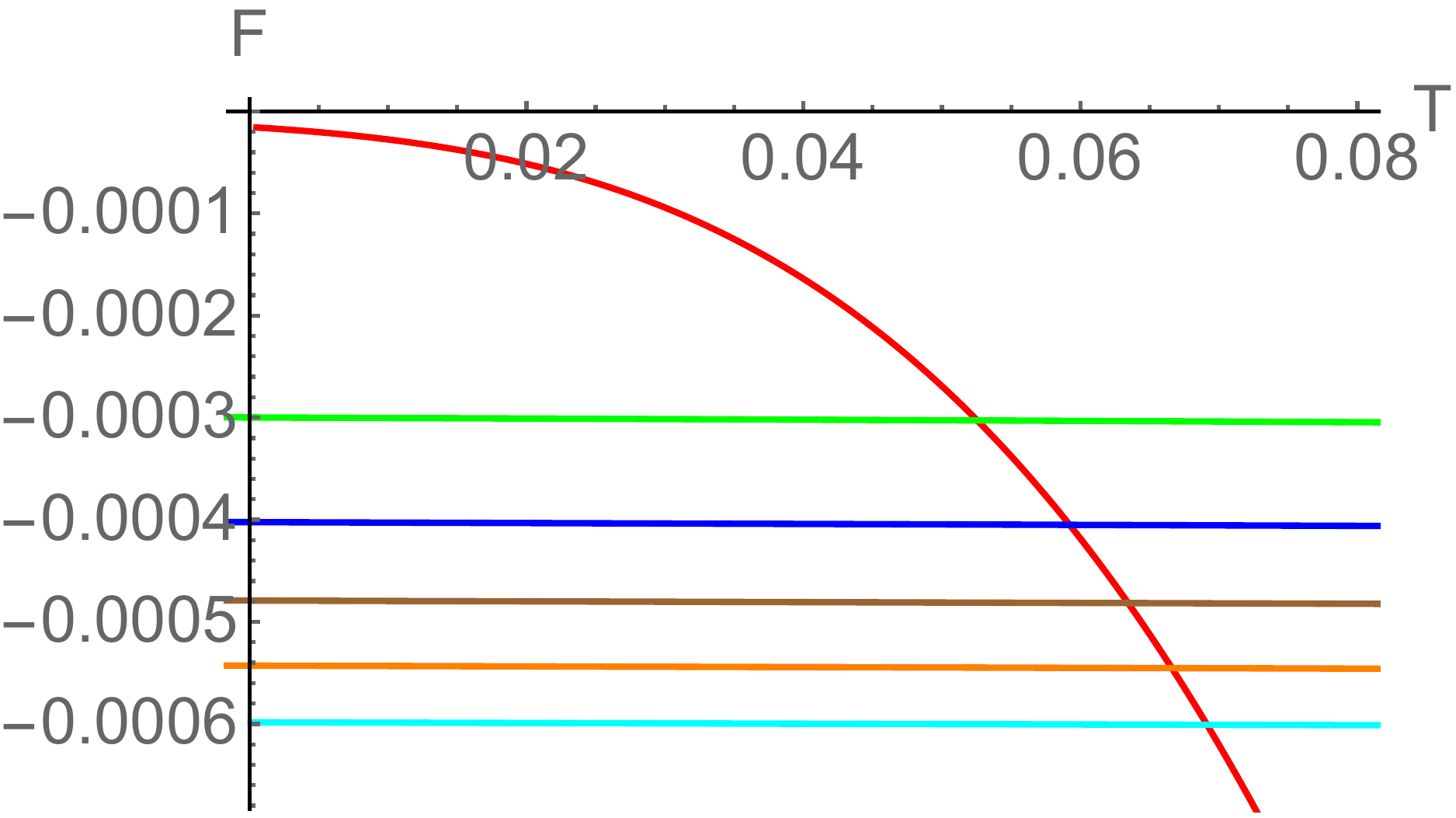}
\caption{ \small  Normalised Gibbs free energy $F$ as a function of Hawking temperature $T$ for various values of $a$.  Here $n=2$, $\mu=0.2$, $\kappa=0$ and $D=4$ are used. Red, green, blue, brown, orange and cyan curves correspond to $a=0$, $0.5$, $1.0$, $1.5$, $2.0$ and $2.5$ respectively. }
\label{TvsFvsamuPt2D4n2}
\end{minipage}
\hspace{0.4cm}
\begin{minipage}[b]{0.5\linewidth}
\centering
\includegraphics[width=2.8in,height=2.3in]{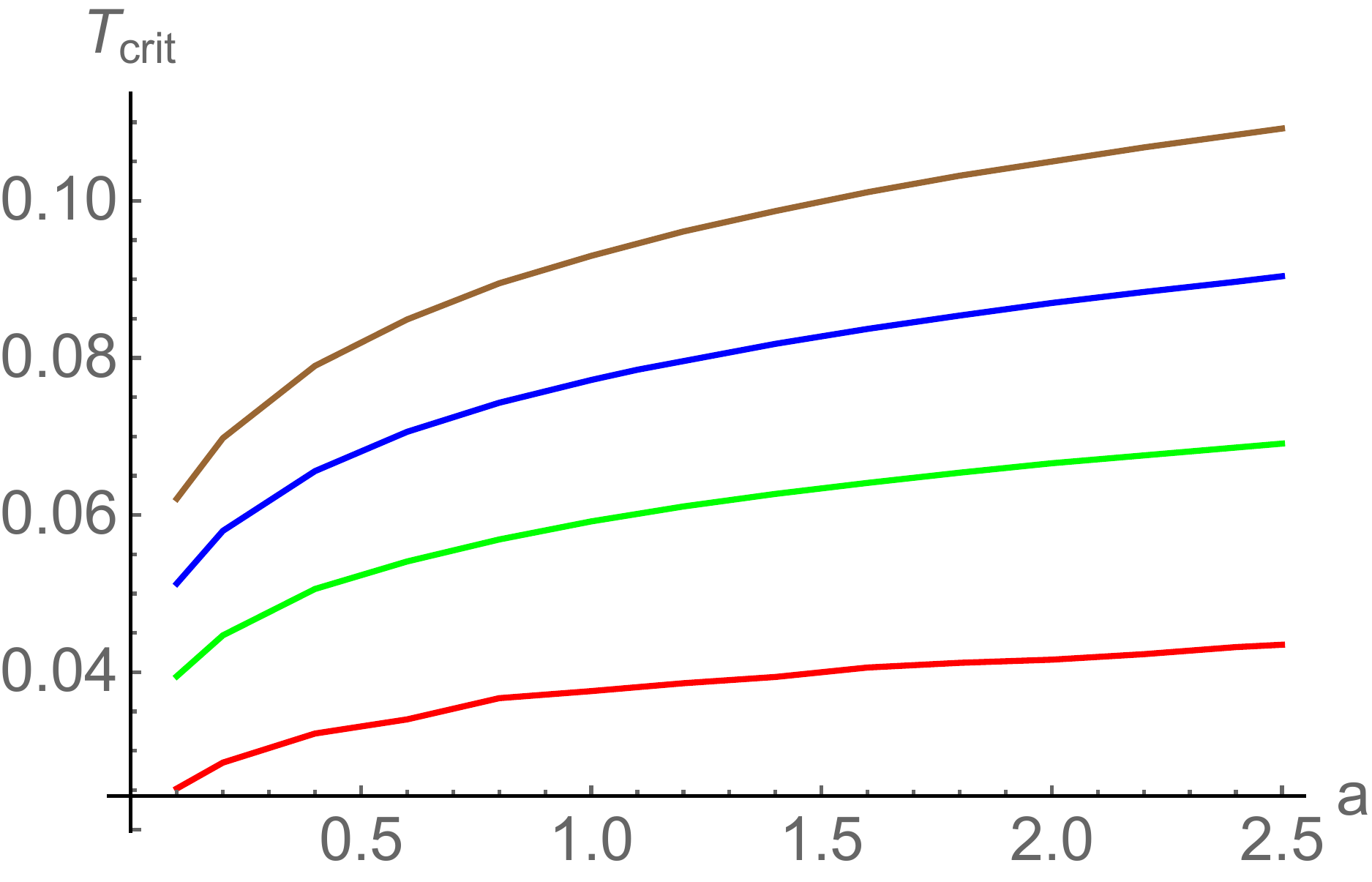}
\caption{\small The hairy/non-hairy black hole phase transition critical temperature $T_{crit}$ as a function $a$ for various values of $\mu$. Here $n=2$, $\kappa=0$ and $D=4$ are used. Red, green, blue and brown curves correspond to $\mu=0.1$, $0.2$, $0.3$ and $0.4$ respectively.}
\label{avsTcritvsmuD4n2}
\end{minipage}
\end{figure}

The free energy analysis further confirms the thermodynamic stability of the charged hairy black hole, especially at low temperatures. Its free energy is again smaller than the planar RN-AdS phase at low temperature. Similarly, the critical temperature for the charged hairy/RN-AdS black hole phase transition depends non-trivially on $a$ and $\mu$. In particular, the temperature range for which hairy black hole configuration remains stable again increases with $a$ and $\mu$. However, just like with $n=1$, the uncharged hairy black holes always have higher free energy than the Schwarzschild AdS black hole. Our results for the thermodynamic stability of the charged planar hairy black hole with $n=2$ are summarised in Figures~\ref{TvsFvsamuPt2D4n2} and \ref{avsTcritvsmuD4n2}.

\begin{figure}[h!]
\begin{minipage}[b]{0.5\linewidth}
\centering
\includegraphics[width=2.8in,height=2.3in]{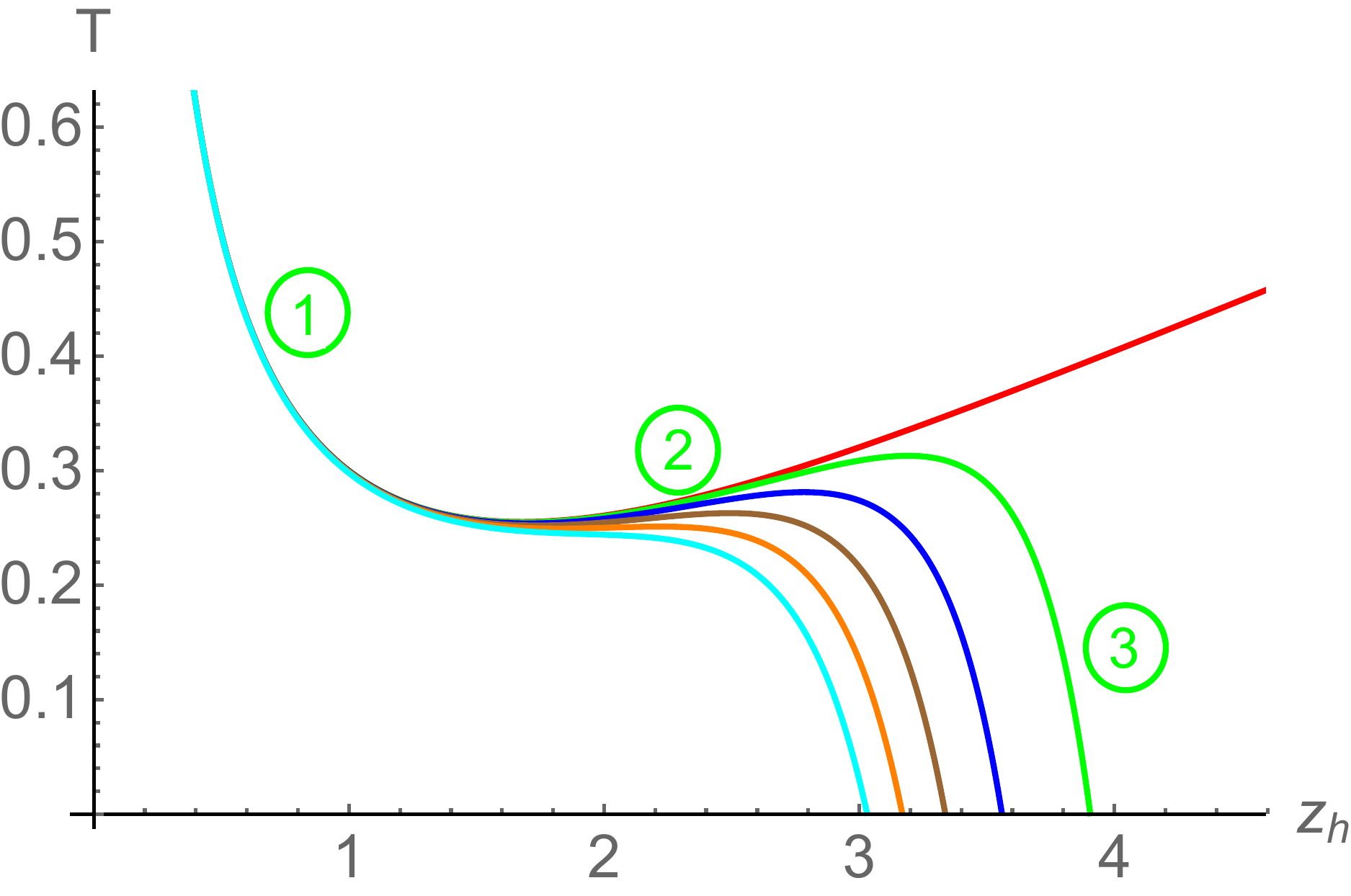}
\caption{ \small Hawking temperature $T$ as a function of horizon radius $z_h$ for various values of chemical potential $\mu$.  Here $n=2$, $a=0.3$, $\kappa=0$ and $D=4$ are used. Red, green, blue, brown, orange and cyan curves correspond to $\mu=0$, $0.05$, $0.1$, $0.15$, $0.2$ and $0.25$ respectively. }
\label{zhvsThvsmuaPt3D4n2}
\end{minipage}
\hspace{0.4cm}
\begin{minipage}[b]{0.5\linewidth}
\centering
\includegraphics[width=2.8in,height=2.3in]{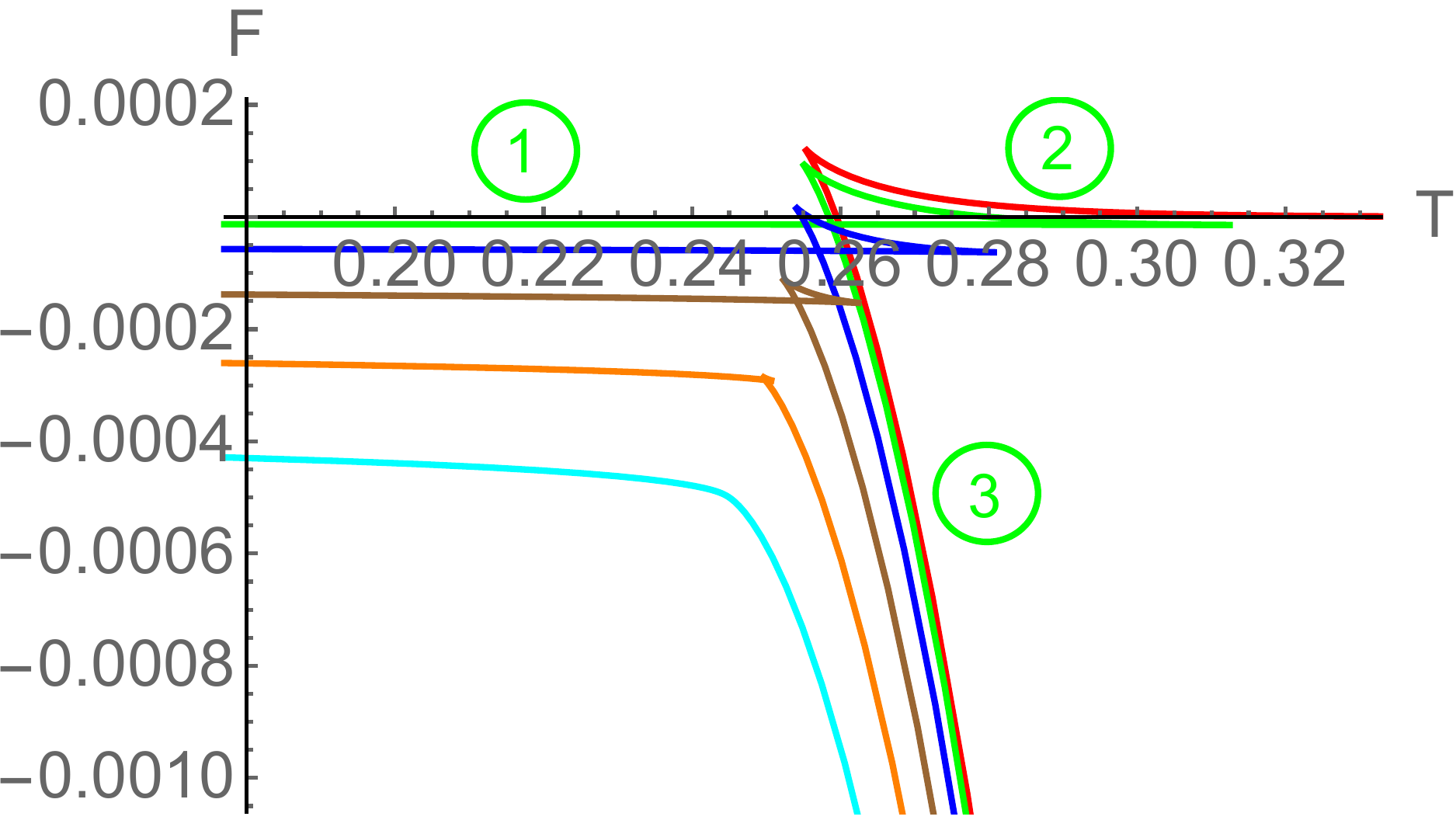}
\caption{\small Normalised Gibbs free energy $F$ as a function of Hawking temperature $T$ for various values of chemical potential $\mu$. Here $n=2$, $a=0.3$, $\kappa=0$ and $D=4$ are used. Red, green, blue, brown, orange and cyan curves correspond to $\mu=0$, $0.05$, $0.1$, $0.15$, $0.2$ and $0.25$ respectively. }
\label{ThvsFvsmuaPt3D4n2}
\end{minipage}
\end{figure}

Actually,  for $n=2$, the thermodynamic structure of the hairy black hole is much more interesting than what is alluded above. In particular, if we consider a hypothetical ensemble of fixed $\mu$ and $a$ then planar hairy black hole can itself undergo various types of phase transitions at high temperatures. The thermodynamic results in the high temperature region are shown in Figures~\ref{zhvsThvsmuaPt3D4n2} and \ref{ThvsFvsmuaPt3D4n2}. For $\mu=0$ (red line), there exists a minimum temperature $T_{min}$ below which no planar hairy black hole solution exists. Moreover, for $T>T_{min}$,  there are now two planar hairy black hole branches in the $(z_h,T)$ plane (instead of one, as in the case of $n=1$). The first branch where $T$ decreases with $z_h$ has the positive specific heat and is stable, whereas the second branch where $T$ increases with $z_h$ has the negative specific heat and is unstable. The stable-unstable nature of these branches can be seen from the free energy behaviour shown in Figure~\ref{ThvsFvsmuaPt3D4n2}. We notice that the free energy of the second black hole branch is not only larger than the first black hole branch but also from the thermal AdS phase, suggesting it to be thermodynamically unfavourable at all temperatures. Importantly, upon varying temperature, the free energy changes its sign at some critical temperature $T_{HP}$, suggesting a Hawking-Page type first-order phase transition between uncharged hairy black hole (first branch) and thermal-AdS at $T_{HP}$.

For small but finite chemical potential $0<\mu<\mu_c$, in addition to a large stable hairy black hole branch (indicated by $\circled{1}$) and an unstable black hole branch (indicated by $\circled{2}$), now a new stable small hairy black hole branch appears at low temperatures (indicated by $\circled{3}$). The intermediate black hole branch, for which the slope in $(z_h,T)$ plane is positive, possess negative specific heat and therefore is unstable, whereas the large and small hairy black hole branches, for which the slope is negative, posses positive specific heat and hence are stable. The hawking temperature now has local maxima and minima and decreases to zero at a finite horizon size. Therefore, at least one black hole phase always exists at all temperature. This behaviour suggests a first-order phase transition from a large hairy black hole to a small hairy black hole as we continuously decrease the temperature. This expectation is confirmed from the free energy behaviour, which exhibits the standard swallowtail like structure - a characteristic feature of the first-order phase transition. The temperature at the kink around which the free energy of large and small hairy black hole phases becomes equal defines the critical temperature $T^{\text{Small/Large}}_{crit}$. Importantly, the free energy is always negative in small and large hairy black hole phases, indicating that these phases have lower free energy than the thermal-AdS and hence are more stable.

\begin{figure}[h!]
\begin{minipage}[b]{0.5\linewidth}
\centering
\includegraphics[width=2.8in,height=2.3in]{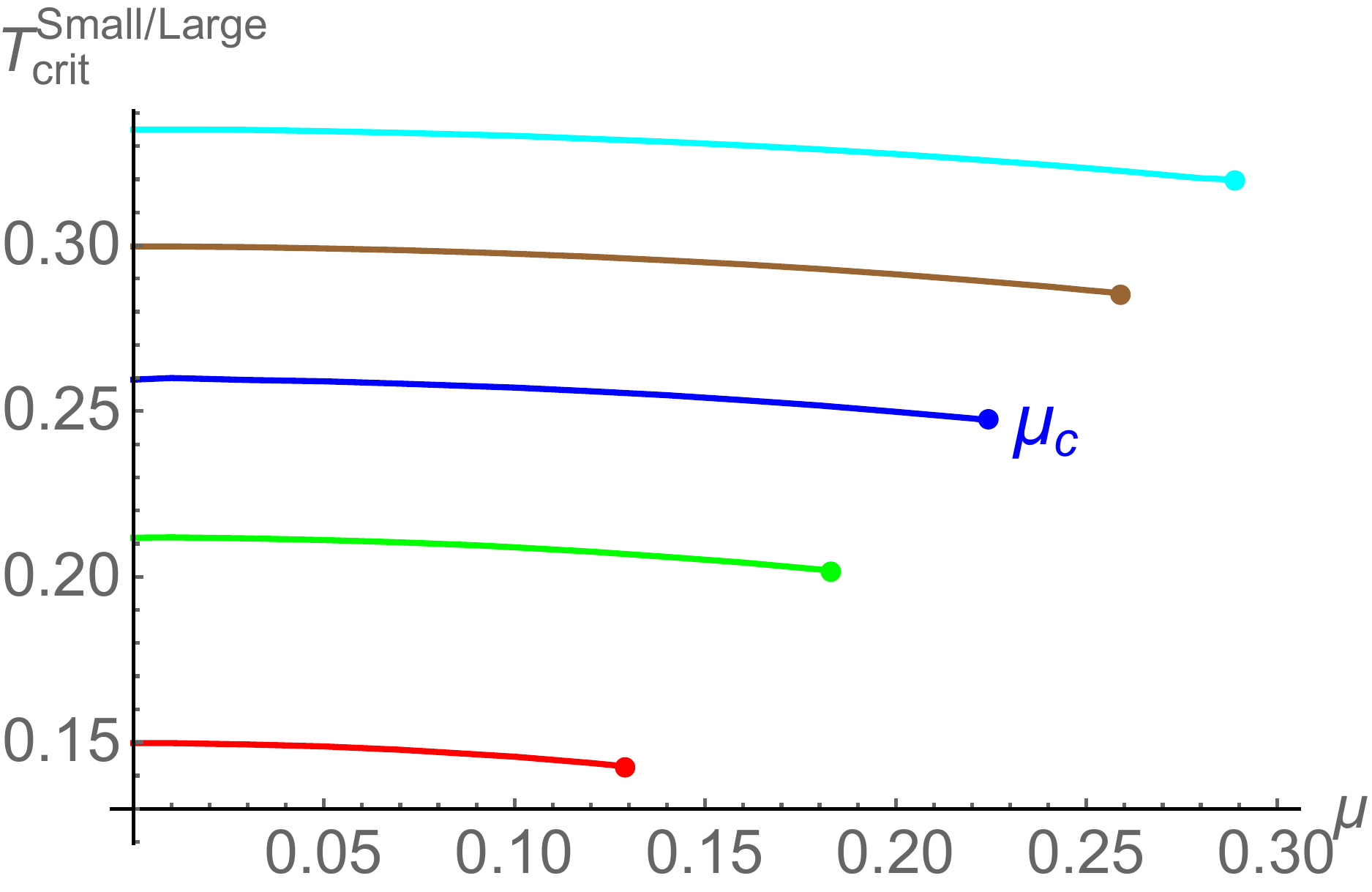}
\caption{ \small The critical temperature $T^{\text{Small/Large}}_{crit}$ as a function $\mu$ for various values of $a$.  Here $n=2$, $\kappa=0$ and $D=4$ are used. Red, green, blue, brown and cyan curves correspond to $a=0.1$, $0.2$, $0.3$, $0.4$ and $0.5$ respectively. }
\label{MuvsTcritvsaD4n2}
\end{minipage}
\hspace{0.4cm}
\begin{minipage}[b]{0.5\linewidth}
\centering
\includegraphics[width=2.8in,height=2.3in]{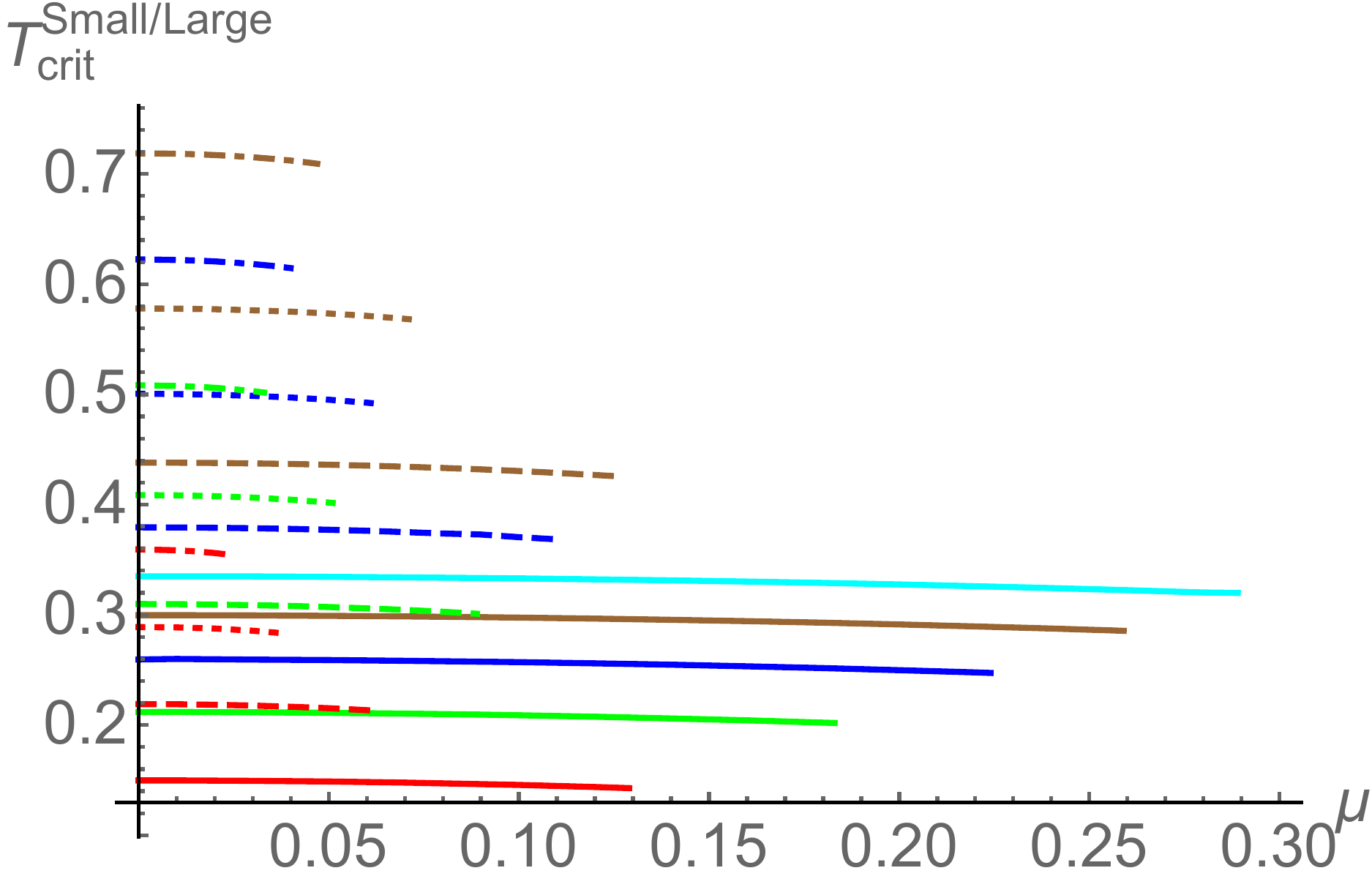}
\caption{\small The critical temperature $T^{\text{Small/Large}}_{crit}$ as a function $\mu$ for various values of $a$ and $D$. Here $n=2$ and $\kappa=0$ are used.  Red, green, blue and brown curves correspond to $a=0.1$, $0.2$, $0.3$ and $0.4$ respectively. Solid, dashed, dotted, and dot-dashed lines correspond to $D=4$, $5$, $6$ and $7$ respectively. }
\label{MuvsTcritvsavsDn2}
\end{minipage}
\end{figure}

As $\mu$ increases, the size of the swallowtail starts shrinking and entirely vanishes at a certain critical chemical potential $\mu_c$. At $\mu_c$, the first-order phase transition between small and large hairy black hole phases stops, and these phases merge together to form a single hairy black hole phase which is stable at all temperatures. The $\mu_c$, therefore, defines a second-order critical point. The complete dependence of $T^{\text{Small/Large}}_{crit}$ on $\mu$ is shown in Figure~\ref{MuvsTcritvsaD4n2}. Overall, the above thermodynamic behaviour in this fixed $\mu$ and $a$ ensemble resembles to the famous Van der Waals type phase transition in black holes \cite{Chamblin:1999tk,Chamblin:1999hg,Dey:2015ytd,Mahapatra:2016dae,Cvetic:1999ne}.

Further, we find the magnitude of $\mu_c$ increases with $a$, suggesting that the hairy small/large black hole phase transition persists for higher and higher values of $\mu$ as $a$ increases. Similar results exist in higher dimensions as well. The difference arises in the magnitude of $\mu_c$, which decreases as $D$ increases for the same value of $a$. The complete phase diagram showing the dependence of $T^{\text{Small/Large}}_{crit}$ on $\mu$, $a$ and $D$ is shown in Figure~\ref{MuvsTcritvsavsDn2}.

It is important to emphasize that the above mentioned Hawking-Page and small/large hairy black hole phase transitions with the planar horizon persist only when $a\neq 0$, \textit{i.e.} when the scalar hair is turned on, and these interesting phase transitions cease to exist for $a=0$. This is consistent with standard results in the literature that planar RN-AdS black holes are always stable and do not undergo any phase transition.

At this point, one might object and say that these interesting Hawking-Page and small/large hairy black hole phase transitions occur at high temperature and therefore are likely to be un-physical, as RN-AdS black hole actually has the lowest free energy at high temperatures. It is important to point out that although the hairy black holes are not thermodynamically favourable at high temperatures, however, they are locally stable. They have positive specific heat and can be in thermal equilibrium with it's surrounding. If we can put the hairy black hole in a large reservoir and can control the parameter $a$, such that it is not allowed to fluctuate, then these Hawking-Page and small/large hairy black hole phase transitions might occur at high temperature. In fact, the occurrence of such phase transitions in the combined system of  AdS black hole and scalar field is essential to many realistic applied AdS/QCD models and have been thoroughly considered in the literature \cite{Dudal:2017max,Dudal:2018ztm,Gursoy:2007er,Gursoy:2008za}.

\subsection{Spherical horizon: $\kappa=1$}
\begin{figure}[ht]
	\subfigure[]{
		\includegraphics[scale=0.4]{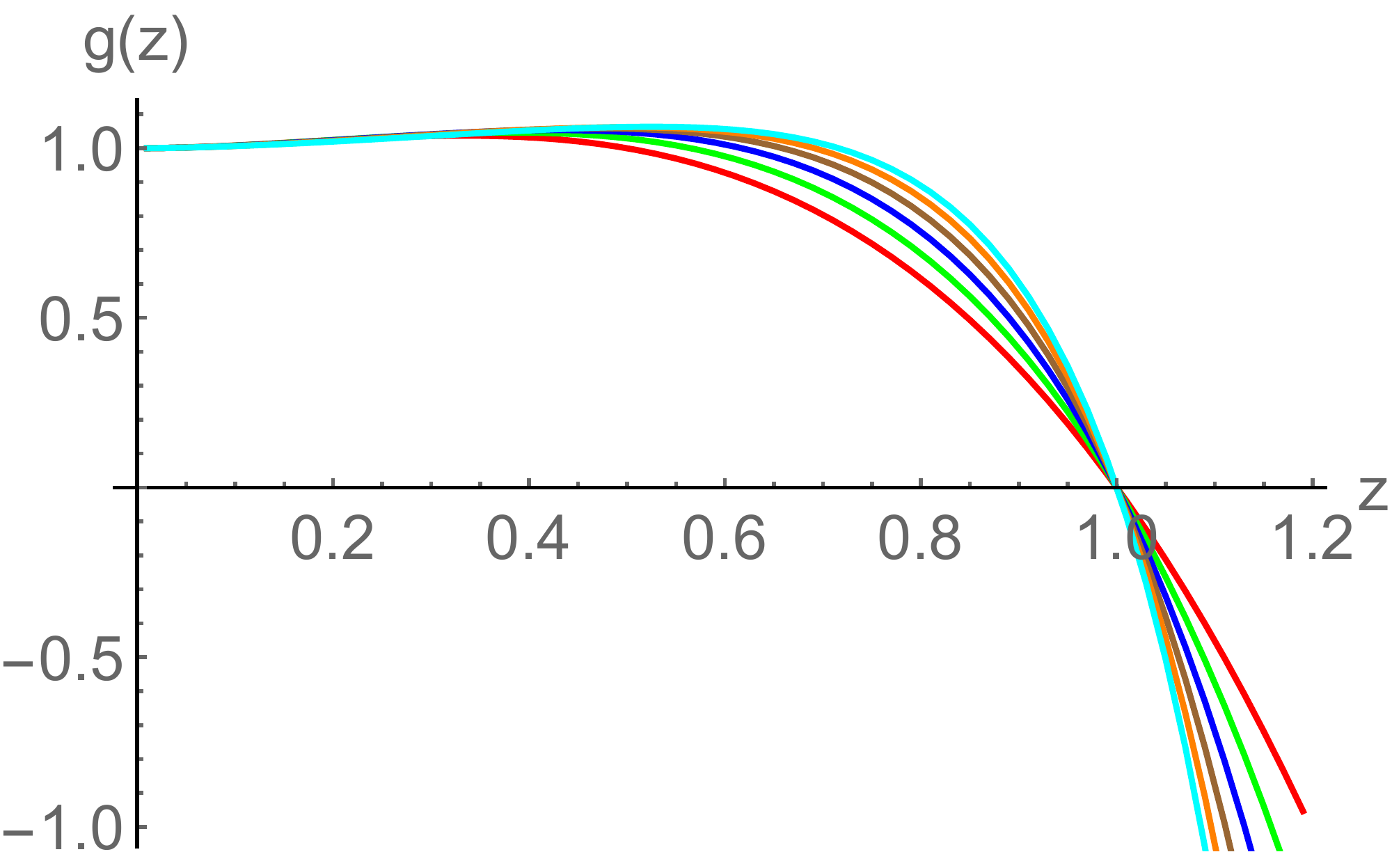}
	}
	\subfigure[]{
		\includegraphics[scale=0.4]{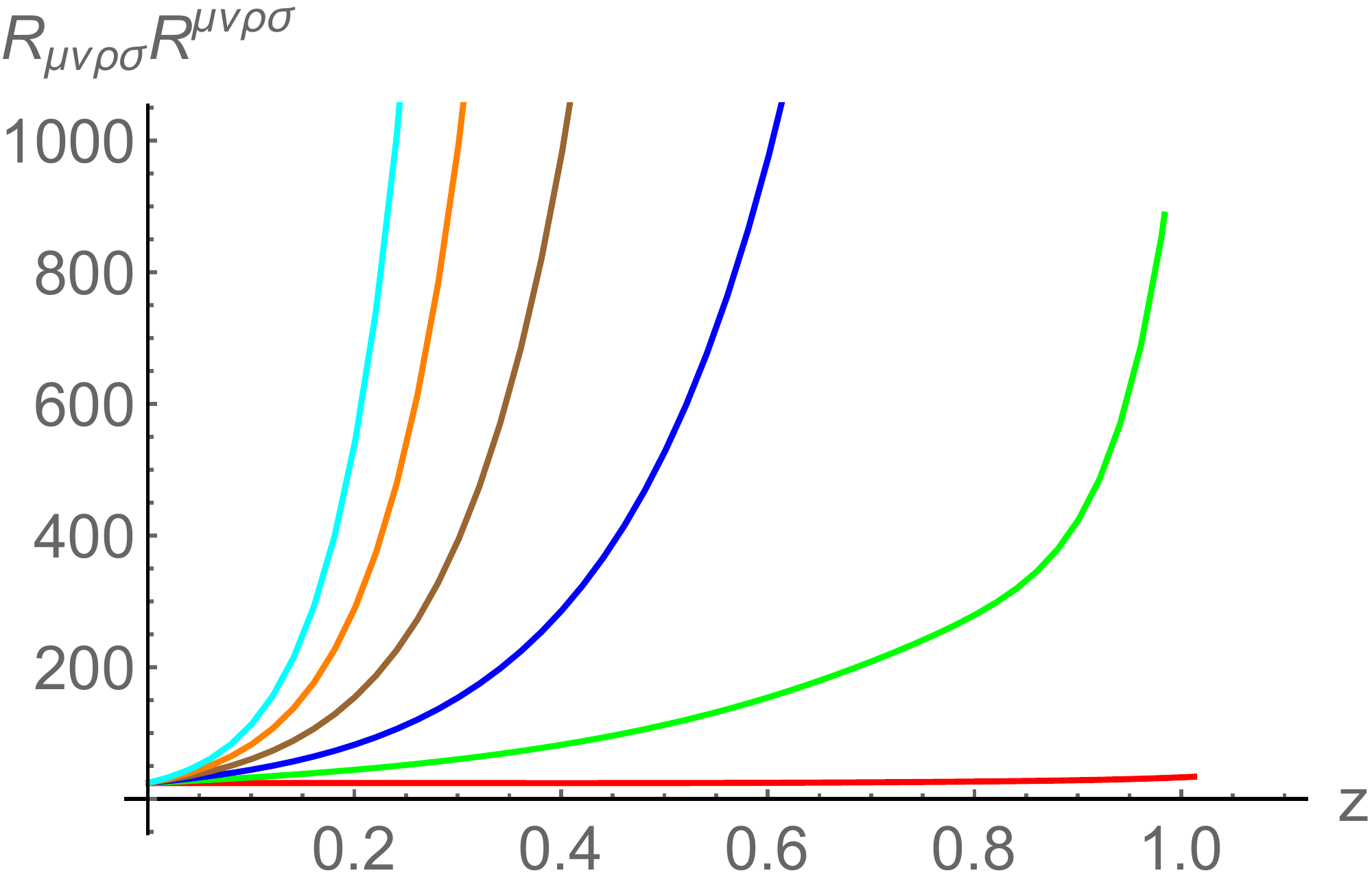}
	}
	\caption{\small The behavior of $g(z)$ and $R_{\mu\nu\rho\lambda}R^{\mu\nu\rho\lambda}$ for different values of $a$. Here $n=1$, $z_h=1$, $\mu=0.1$, $\kappa=1$ and $D=4$ are used. Red, green, blue, brown, orange and cyan curves correspond to $a=0$, $0.5$, $1.0$, $1.5$, $2.0$ and $2.5$ respectively.}
	\label{zvsgvsamuPt1zh1D4n1sph}
\end{figure}

Next, we turn our attention to the spherical horizon. Here we mainly concentrate on $n=1$, as the results with $n=2$ are qualitatively similar. With $A(z)=-az$ and $\kappa=1$, the expression of $g(z)$ in four dimensions reduces to,
\begin{eqnarray}
& & \frac{\pi}{2a^2} \left[ \mathcal{G}_{3,4}^{2,2}\left(-2
   a z\left|
\begin{array}{c}
 1,3,\frac{7}{2} \\
 3,3,0,\frac{7}{2} \\
\end{array}
\right.\right)  -  \frac{e^{2 a z} \left(2 a^2 z^2-2 a z+1\right)-1}{e^{2
   a z_h} \left(2 a^2 z_h^2-2 a z_h+1\right)-1} \mathcal{G}_{3,4}^{2,2}\left(-2
   a z_h \left|
\begin{array}{c}
 1,3,\frac{7}{2} \\
 3,3,0,\frac{7}{2} \\
\end{array}
\right.\right) \right] + \nonumber \\
& &  \frac{\mu ^2 \left[\left(z-z_h\right) e^{2 a \left(z_h+z\right)} \left[z^2 \left(2 a z_h
   \left(a z_h-1\right)+1\right)+z z_h \left(1-2 a z_h\right)+z_h^2\right]+z_h^3 e^{2 a
   z_h}-z^3 e^{2 a z}\right]}{2 a z_h^2 \left[e^{2 a z_h} \left(2 a z_h
   \left(a z_h-1\right)+1\right)-1\right]} \nonumber \\
&&  + 1 + z^2 - \frac{e^{2 a z} \left(2 a^2 z^2-2 a z+1\right)-1}{e^{2
   a z_h} \left(2 a^2 z_h^2-2 a z_h+1\right)-1}(1+z_h^2)
\label{gD4n1spherical}
\end{eqnarray}
where $\mathcal{G}$ is the Meijer function \cite{wolfram,Beals}. One can explicitly check that $g(z)$ reduces to the standard spherical RN-AdS expression in the limit $a\rightarrow 0$. The expression for $\phi(z)$ remains the same as in eq.~(\ref{phiD4n1planar}). Therefore, the scalar field remains finite and regular everywhere in the bulk. The behaviour of $g(z)$ and Kretschmann scalar are shown in Figure~\ref{zvsgvsamuPt1zh1D4n1sph}. We again find a smooth behaviour of the Kretschmann scalar, suggesting a well-behaved nature of the spacetime geometry.

\begin{figure}[h!]
\begin{minipage}[b]{0.5\linewidth}
\centering
\includegraphics[width=2.8in,height=2.3in]{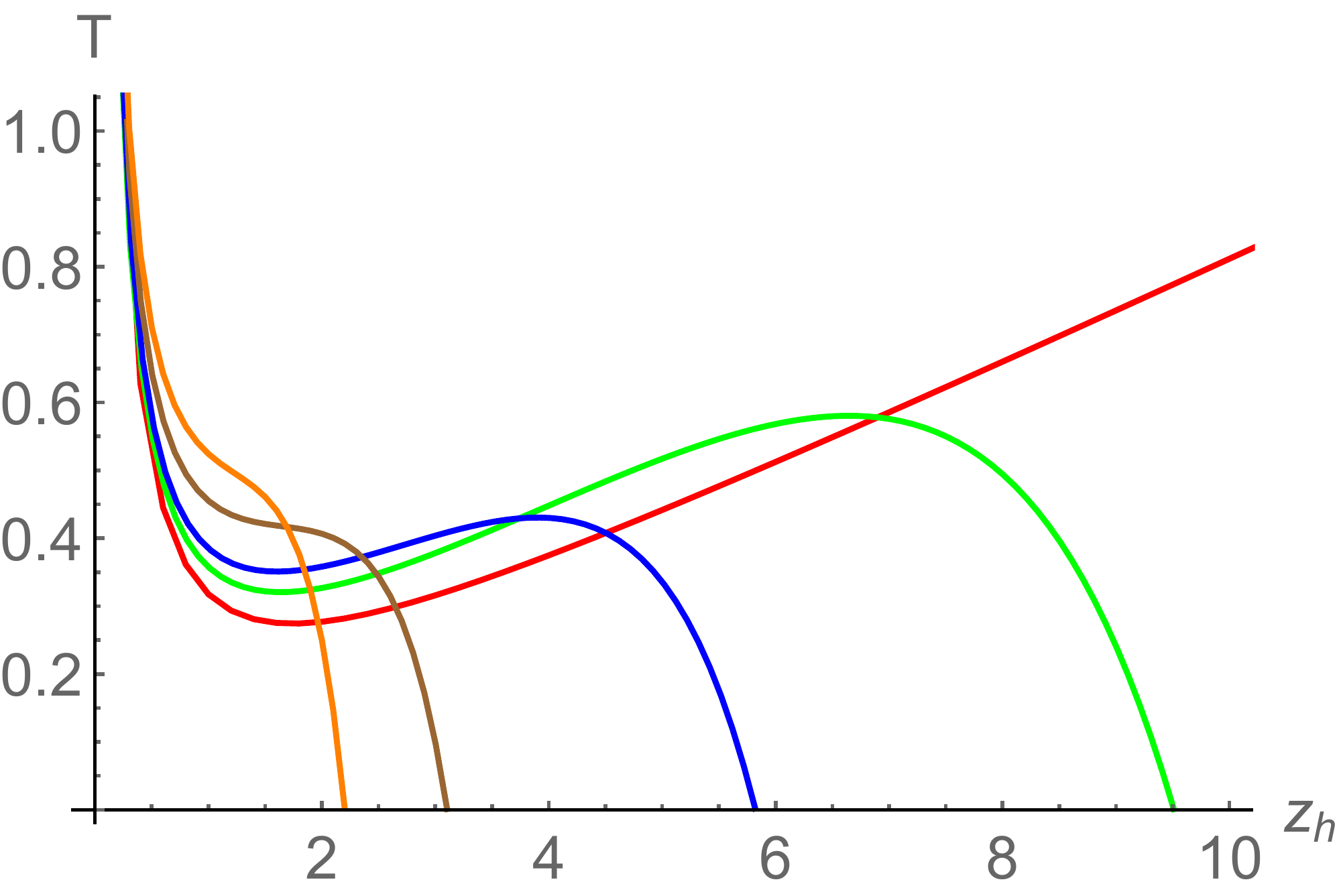}
\caption{ \small Hawking temperature $T$ as a function of horizon radius $z_h$ for various values of $a$.  Here $n=1$, $\mu=0.2$, $\kappa=1$ and $D=4$ are used. Red, green, blue, brown and orange curves correspond to $a=0$, $0.3$, $0.5$, $1.0$ and $1.5$ respectively. }
\label{zhvsThvsamuPt2D4n1sph}
\end{minipage}
\hspace{0.4cm}
\begin{minipage}[b]{0.5\linewidth}
\centering
\includegraphics[width=2.8in,height=2.3in]{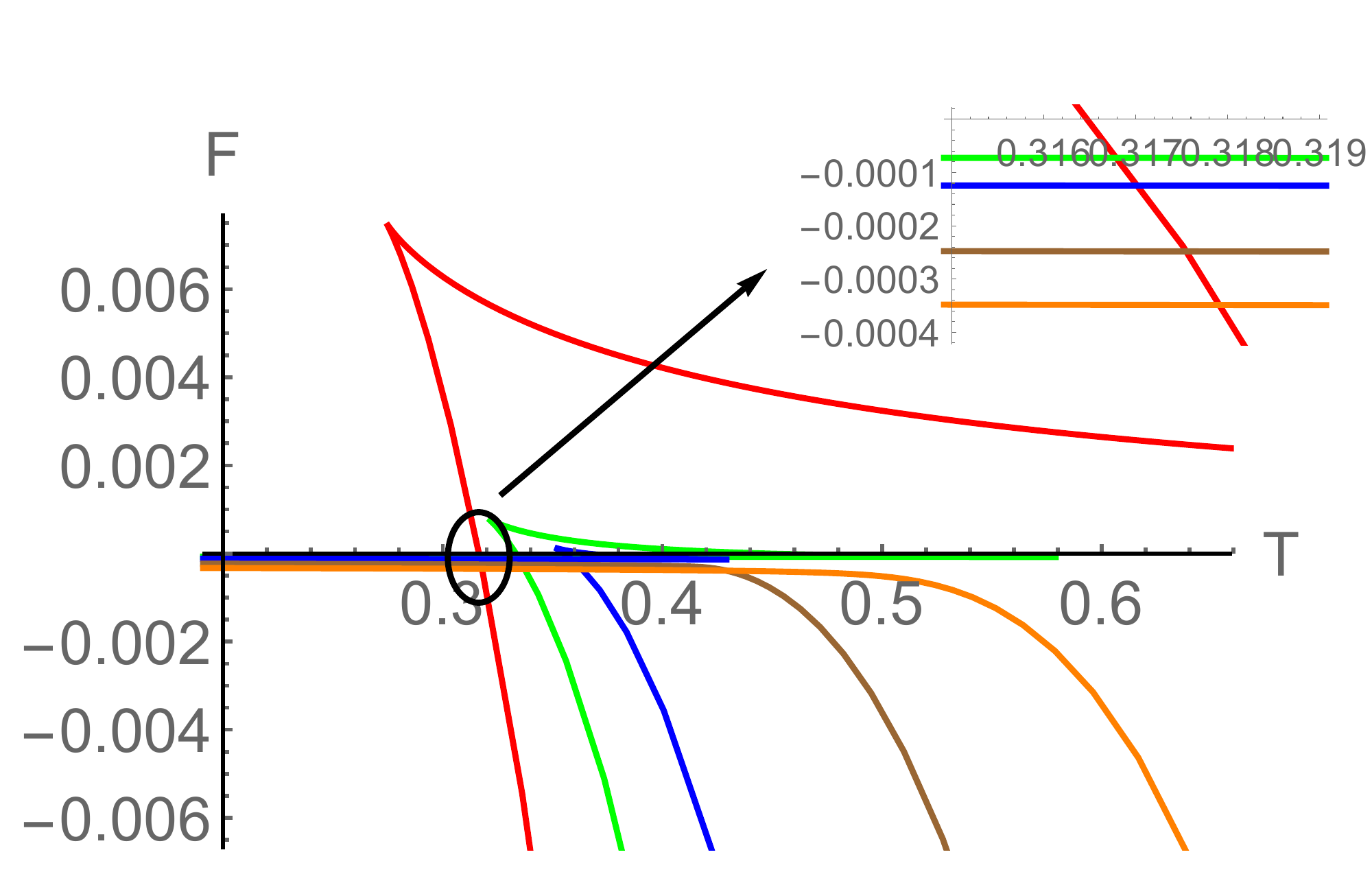}
\caption{\small Normalised Gibbs free energy $F$ as a function of Hawking temperature $T$ for various values of $a$. Here $n=1$, $\mu=0.2$, $\kappa=1$ and $D=4$ are used. Red, green, blue, brown and orange curves correspond to $a=0$, $0.3$, $0.5$, $1.0$ and $1.5$ respectively. }
\label{TvsFvsamuPt2D4n1sph}
\end{minipage}
\end{figure}

Next, we analyse the thermodynamic structure of the charged hairy spherical black hole. Our results are shown in Figures~\ref{zhvsThvsamuPt2D4n1sph} and \ref{TvsFvsamuPt2D4n1sph}. We find drastic changes in the thermodynamic structure of the black hole in the presence of scalar hair, especially in the regime of small potential. As is well known, in the regime of small potential $\mu<\mu_c=2$, RN-AdS black hole has two branches, a stable branch with larger radii (smaller $z_h$) and an unstable branch with smaller radii (larger $z_h$). Importantly, for $\mu<\mu_c$, RN-AdS black hole exists only above a certain minimum temperature. This is shown by a red line in Figure~\ref{zhvsThvsamuPt2D4n1sph}. However, with scalar hair, the charged black hole now exists at all temperature. Depending upon the magnitude of $a$, there can now be one or three hairy black hole branches. In particular, for large $a$ there is one stable branch whereas for small $a$ there are two stable and one unstable branches. Therefore, for the hairy case, at least one black hole branch is always stable and exists at all temperatures. Since the RN-AdS black hole phase exists only above a minimum temperature $T_{min}$, the charged hairy black hole phase therefore naturally becomes more favourable below $T_{min}$ \footnote{There exists a Hawking/Page phase transition at $T_{crit}>T_{min}$. Therefore, in principle,  RN-AdS black hole is thermodynamically favoured only above $T_{crit}$.}.

The charged hairy phase can continue to dominate the thermodynamics at $T>T_{min}$ as well. In particular, with higher $a$, its free energy can remain smaller than the RN-AdS phase for larger $T$ (see Figure~\ref{TvsFvsamuPt2D4n1sph}).  Moreover, the free energy of the stable hairy black hole branches is always negative, further indicating thermodynamically favoured nature of these charged hairy black holes. Importantly, as in the case of planar horizon, the spherical hairy black hole becomes favourable only when $\mu\neq0$. For $\mu=0$, the free energy of spherical RN-AdS black hole is always smaller than the spherical hairy black hole.

As in the previous section, if one again considers a hypothetical ensemble of fixed finite $a$ and $\mu$($<\mu_c$), then there can occur small/large charged hairy black hole transitions, especially at high temperatures. The corresponding critical temperature again non-trivially depends on $a$ and $\mu$, in particular, it increases(decreases) as $a$($\mu$) increases. The critical temperature phase diagram is qualitatively similar to Figure~\ref{MuvsTcritvsaD4n2}.

\begin{figure}[h!]
\begin{minipage}[b]{0.5\linewidth}
\centering
\includegraphics[width=2.8in,height=2.3in]{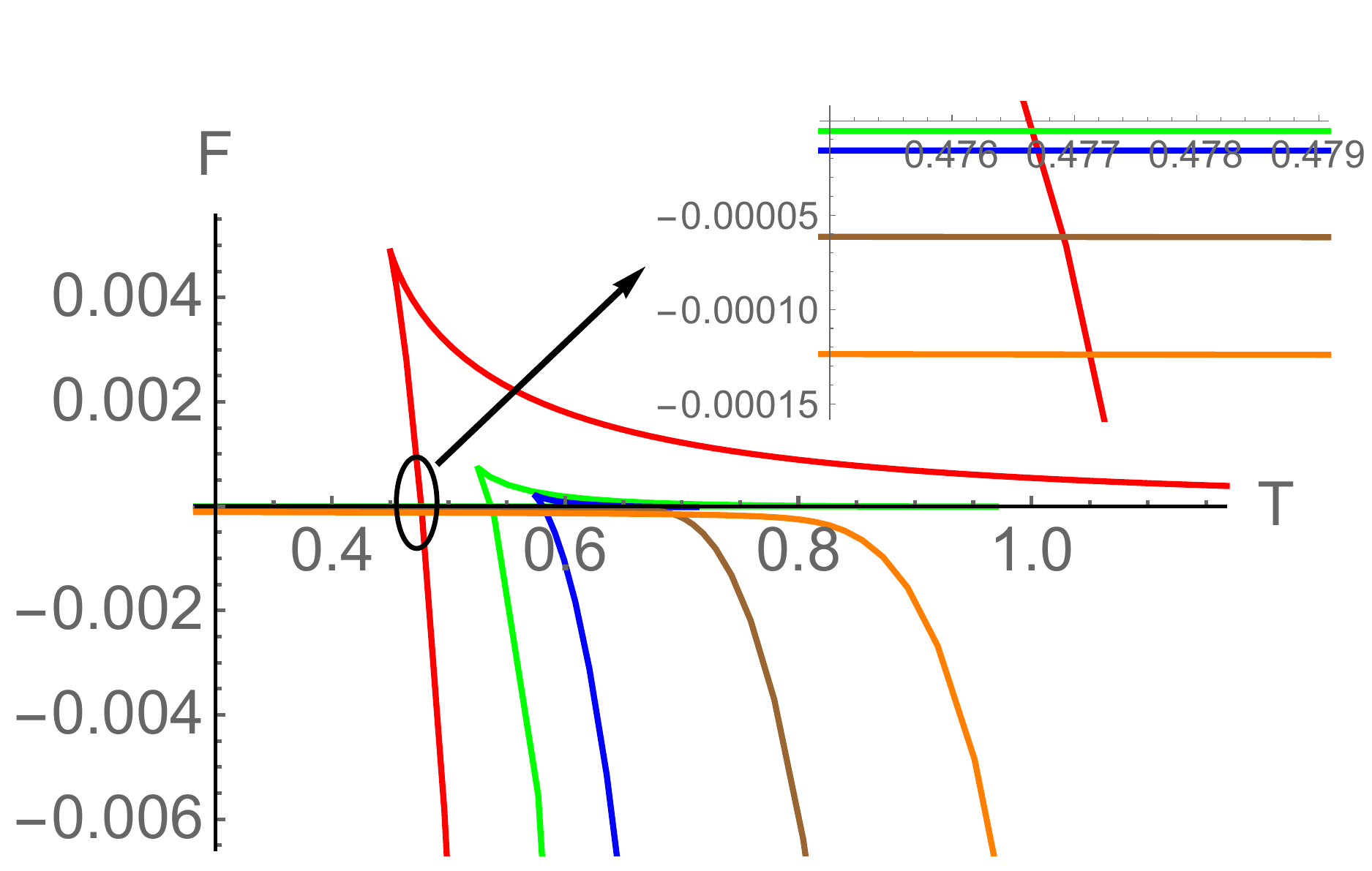}
\caption{ \small Normalised Gibbs free energy $F$ as a function of Hawking temperature $T$ for various values of $a$. Here $n=1$, $\mu=0.1$, $\kappa=1$ and $D=5$ are used. Red, green, blue, brown and orange curves correspond to $a=0$, $0.3$, $0.5$, $1.0$ and $1.5$ respectively. }
\label{TvsFvsamuPt1D5n1sph}
\end{minipage}
\hspace{0.4cm}
\begin{minipage}[b]{0.5\linewidth}
\centering
\includegraphics[width=2.8in,height=2.3in]{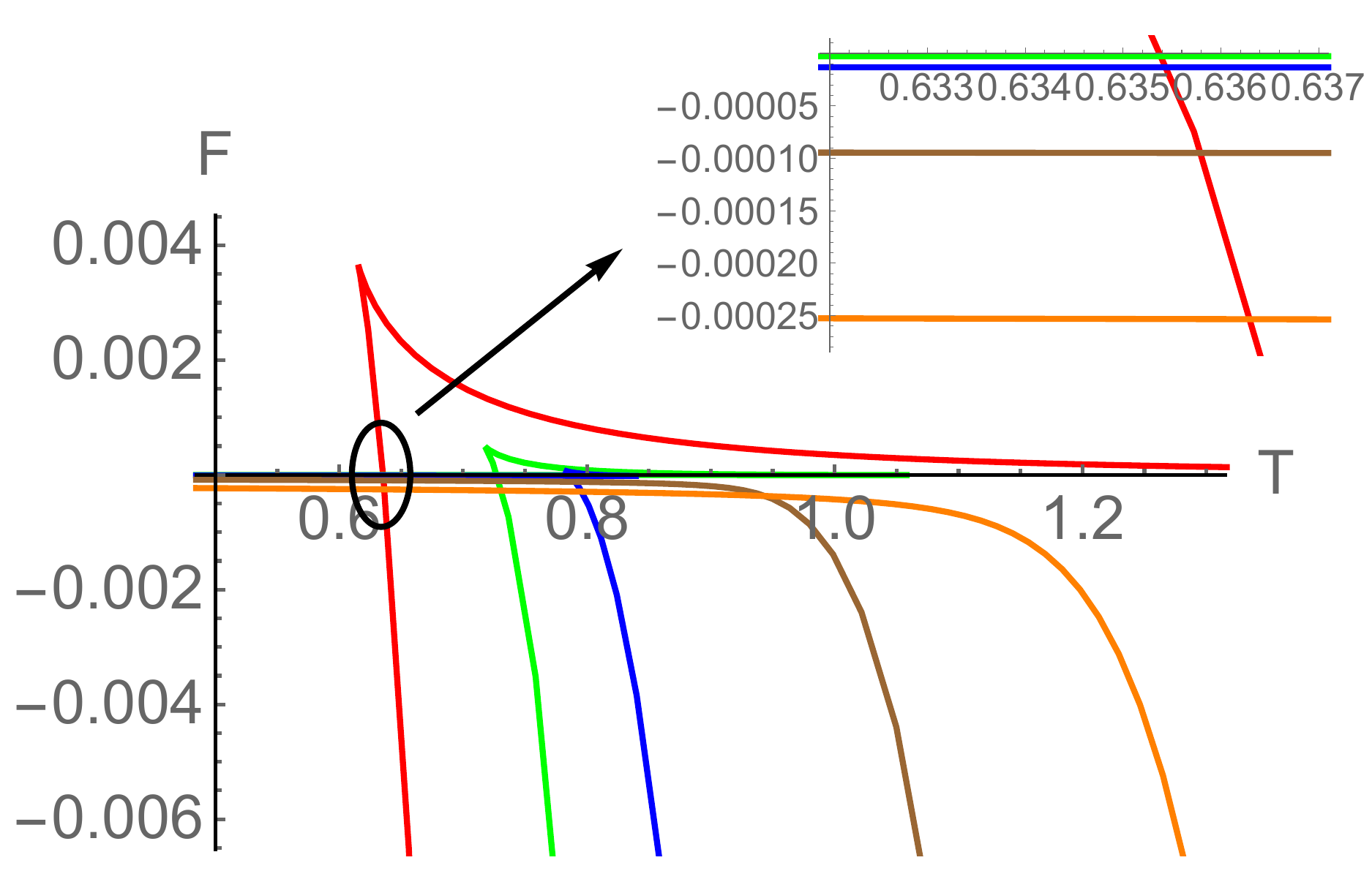}
\caption{\small Normalised Gibbs free energy $F$ as a function of Hawking temperature $T$ for various values of $a$. Here $n=1$, $\mu=0.1$, $\kappa=1$ and $D=6$ are used. Red, green, blue, brown and orange curves correspond to $a=0$, $0.3$, $0.5$, $1.0$ and $1.5$ respectively. }
\label{TvsFvsamuPt1D6n1sph}
\end{minipage}
\end{figure}

\begin{figure}[h!]
\begin{minipage}[b]{0.5\linewidth}
\centering
\includegraphics[width=2.8in,height=2.3in]{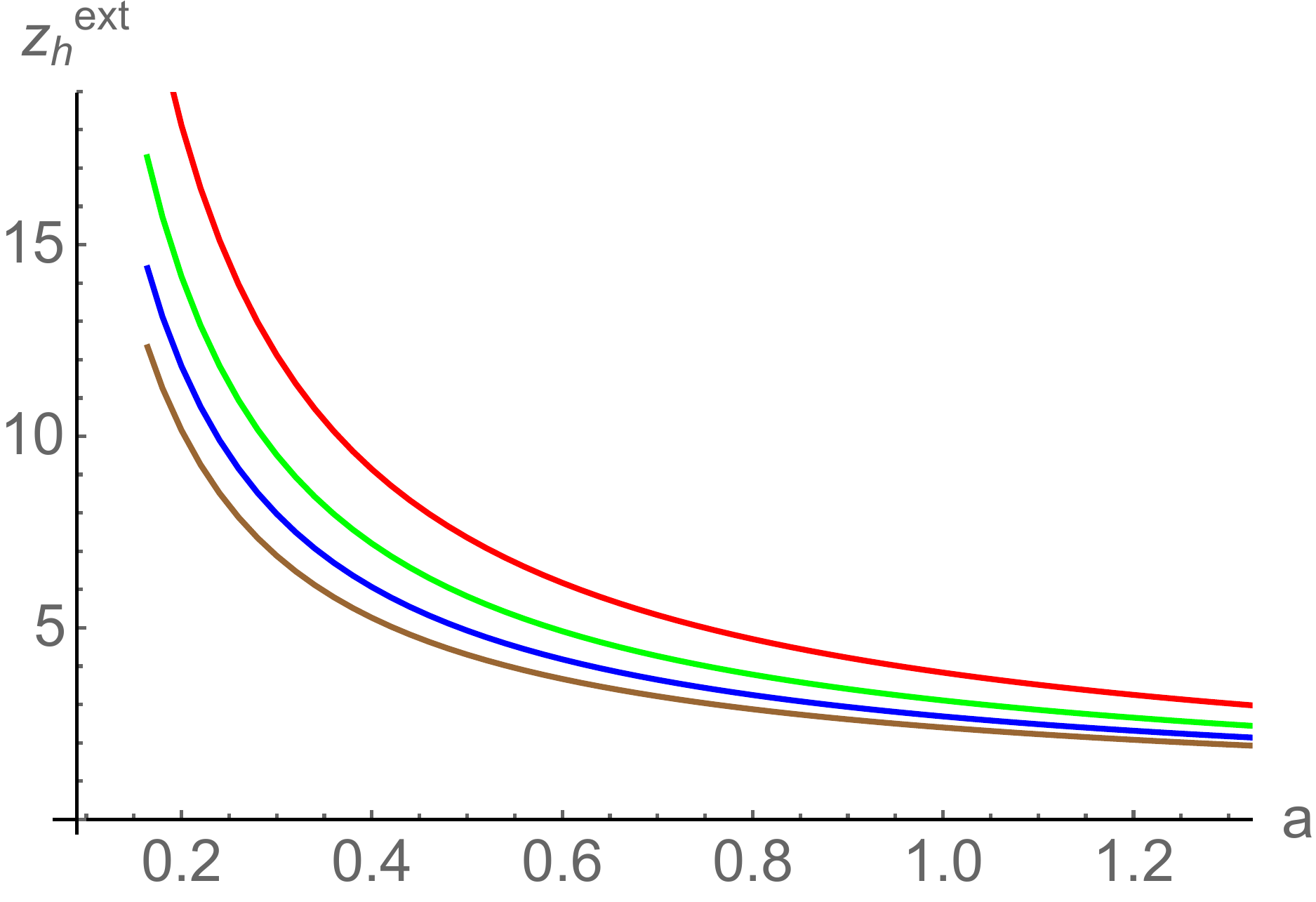}
\caption{ \small The variation of extremal black hole horizon radius $z_{h}^{\text{ext}}$ as function of $a$. Here $n=1$, $\kappa=1$ and $D=4$ are used. Red, green, blue and brown curves correspond to $\mu=0.1$, $0.2$, $0.3$ and $0.4$ respectively.}
\label{avsZhextvsmuD4n1sph}
\end{minipage}
\hspace{0.4cm}
\begin{minipage}[b]{0.5\linewidth}
\centering
\includegraphics[width=2.8in,height=2.3in]{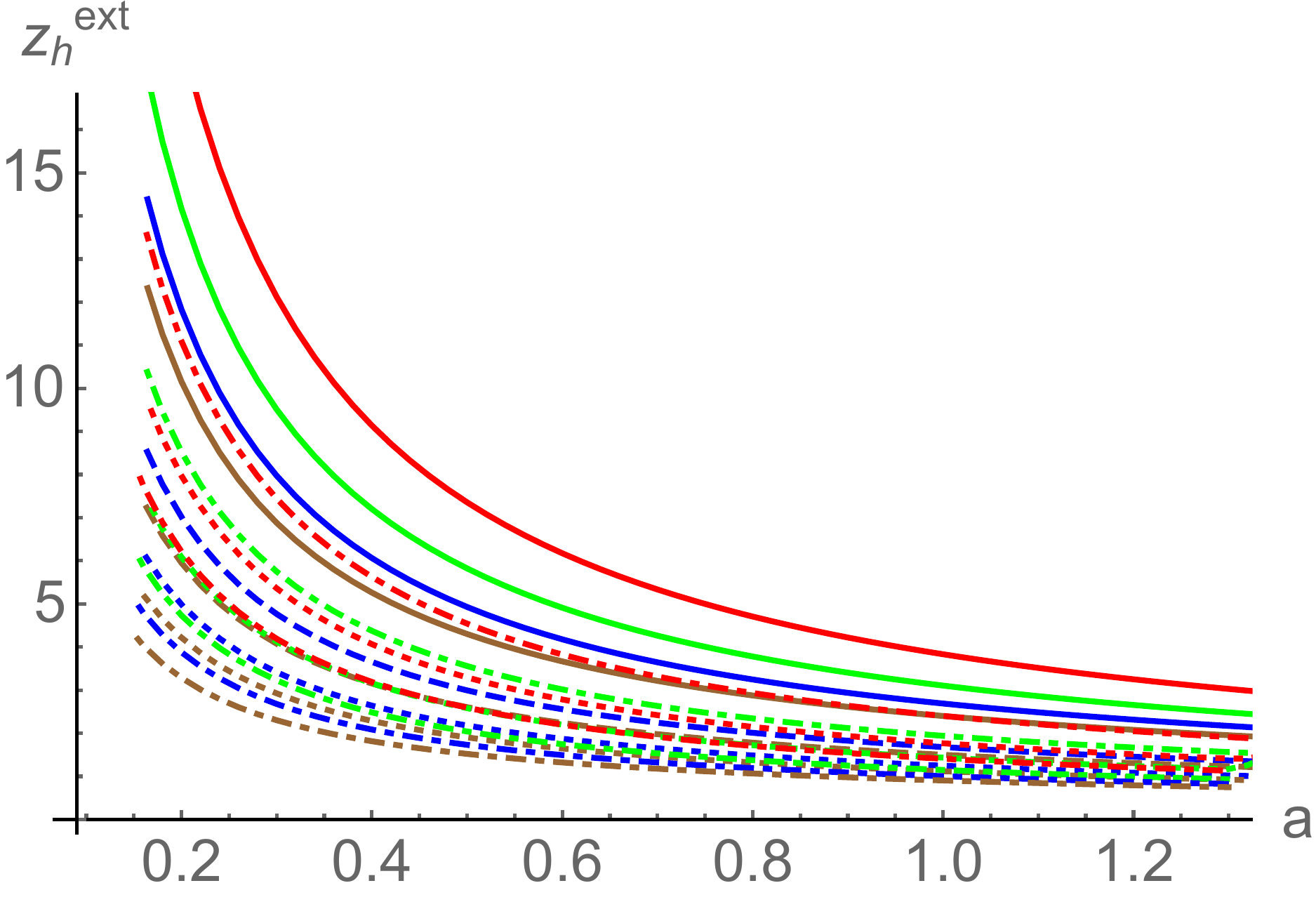}
\caption{\small The variation of extremal black hole horizon radius $z_{h}^{\text{ext}}$ as function of $a$. Here $n=1$ and $\kappa=1$ are used. Red, green, blue and brown curves correspond to $\mu=0.1$, $0.2$, $0.3$ and $0.4$ respectively. Solid, dashed, dotted and dotdashed curves correspond to $D=4$, $5$, $6$ and $7$ respectively.}
\label{avsZhextvsmuvsDn1sph}
\end{minipage}
\end{figure}

On the other hand, for higher potential $\mu>\mu_c$, there exists only one stable black hole branch in the RN-AdS black hole. The same result holds in the case of hairy black holes as well. The difference arises in the magnitude of $\mu_c$, which decreases as $a$ increases. Interestingly, irrespective of whether $\mu>\mu_c$ or not (provided $\mu\neq0$), the charged hairy black hole can become extremal (as opposed to the RN-AdS black hole, which becomes extremal only when $\mu>\mu_c$). Moreover, for a fixed $\mu$, the hairy black hole becomes extremal for smaller $z_h$ as $a$ increases. This indicates that the size of the extremal hairy black hole increases as the strength of the scalar hair increases. The dependence of the extremal horizon radius $z_{h}^{\text{ext}}$ on $a$ and $\mu$ is shown in Figure~\ref{avsZhextvsmuD4n1sph}.

Further, we analyse the hairy black hole thermodynamics in higher dimensions. Our results for the free energy in $D=5$ and $D=6$ are shown in Figures~\ref{TvsFvsamuPt1D5n1sph} and \ref{TvsFvsamuPt1D6n1sph}. We again find thermodynamically favoured charged hairy black holes at low temperatures, with the hairy/RN-AdS critical temperature to be increasing with $a$. Importantly, even with higher $D$, the hairy black hole becomes favourable only when $\mu\neq0$ whereas the free energy of RN-AdS black hole is always smaller when $\mu=0$. Moreover, similar to the case when $D=4$, the charged hairy black holes in higher dimensions can become extremal as well. The corresponding extremal horizon radius $z_{h}^{\text{ext}}$ again depends non-trivially on $a$ and $\mu$.  The complete phase diagram showing the dependence of $z_{h}^{\text{ext}}$ on $a$ and $\mu$ in different dimensions is shown in Figure~\ref{avsZhextvsmuvsDn1sph}. Notice that, for a fixed value of $a$ and $\mu$, as spacetime dimension increases the black hole becomes extremal at smaller $z_h$ (and hence at larger sizes).

\subsection{Hyperbolic horizon: $\kappa=-1$}

\begin{figure}[ht]
	\subfigure[]{
		\includegraphics[scale=0.4]{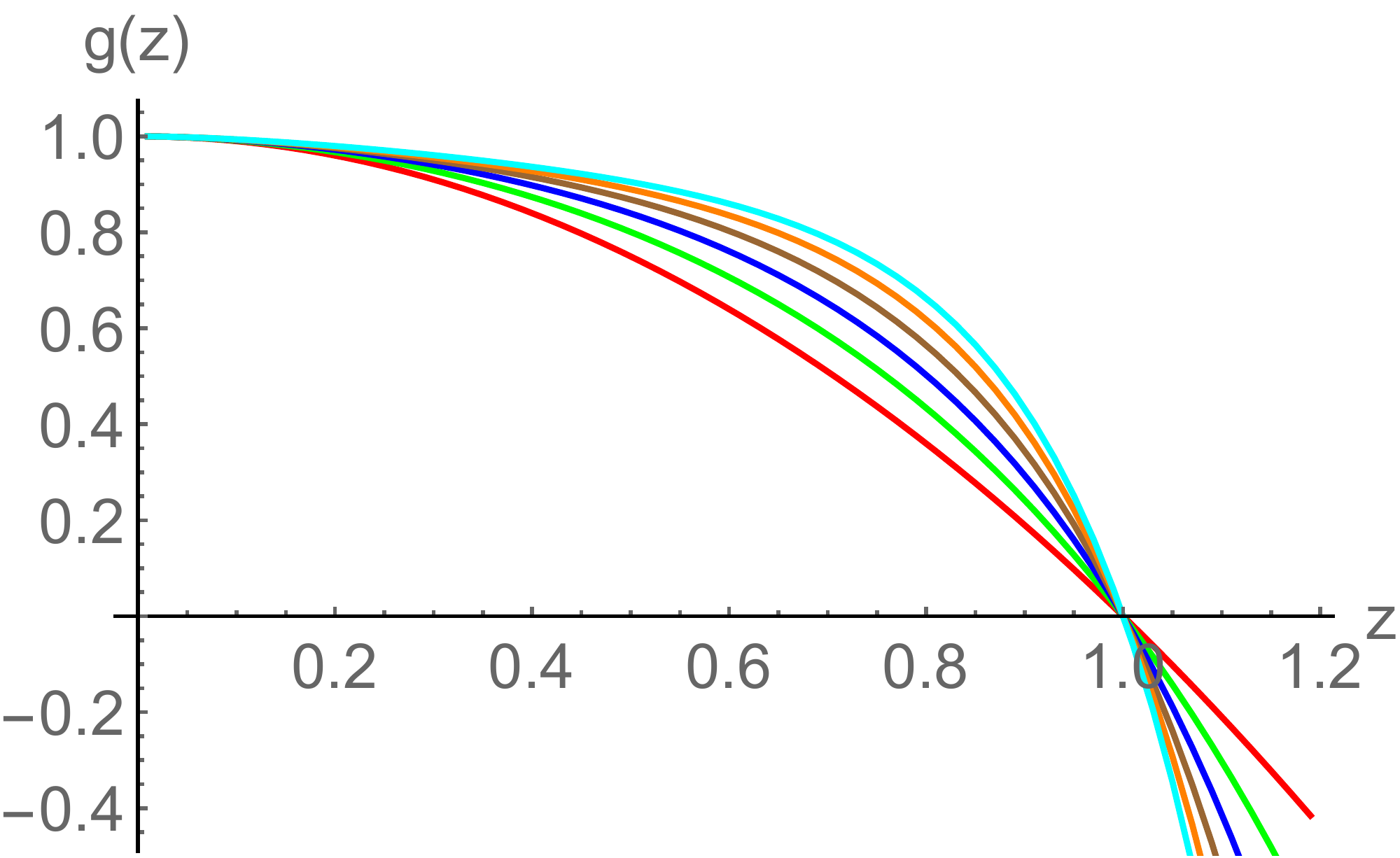}
	}
	\subfigure[]{
		\includegraphics[scale=0.4]{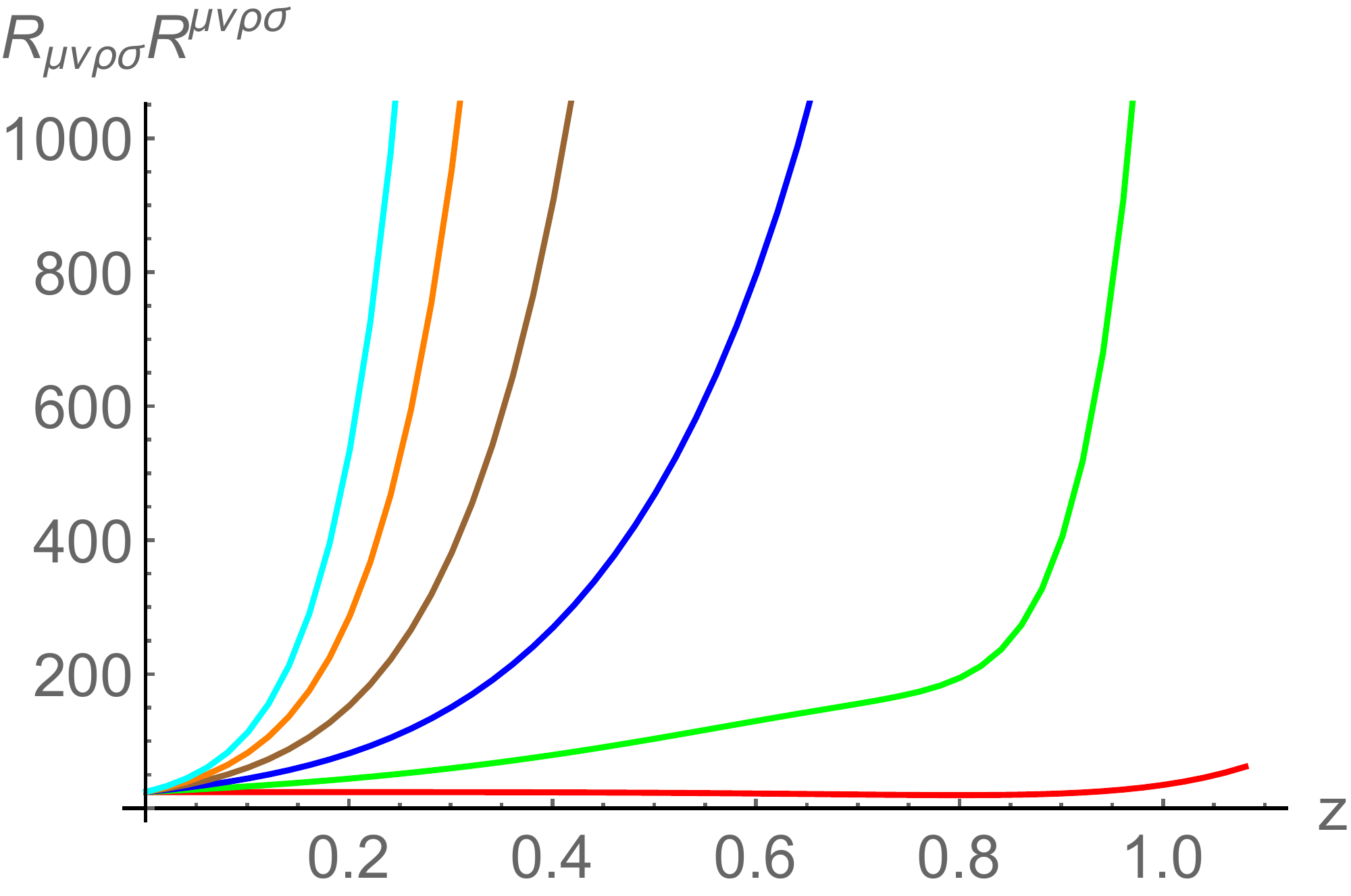}
	}
	\caption{\small The behavior of $g(z)$ and $R_{\mu\nu\rho\lambda}R^{\mu\nu\rho\lambda}$ for different values of $a$. Here $n=1$, $z_h=1$, $\mu=0.1$, $\kappa=-1$ and $D=4$ are used. Red, green, blue, brown, orange and cyan curves correspond to $a=0$, $0.5$, $1.0$, $1.5$, $2.0$ and $2.5$ respectively.}
	\label{zvsgvsamuPt1zh1D4n1hyp}
\end{figure}
Let us now briefly discuss the results with the hyperbolic horizon. Here we present results for $n=1$ only, as the results with $n=2$ are qualitatively similar. Analytic expression of $g(z)$ can again be obtained. In Figures~\ref{zvsgvsamuPt1zh1D4n1hyp}, the behaviour of $g(z)$ and Kretschmann scalar is shown. Similar to the case of planar and spherical horizons, once again we find a well-behaved hairy geometry having no divergences in the bulk spacetime. Similarly, the scalar field remains finite and regular everywhere in the bulk as well.

\begin{figure}[h!]
\begin{minipage}[b]{0.5\linewidth}
\centering
\includegraphics[width=2.8in,height=2.3in]{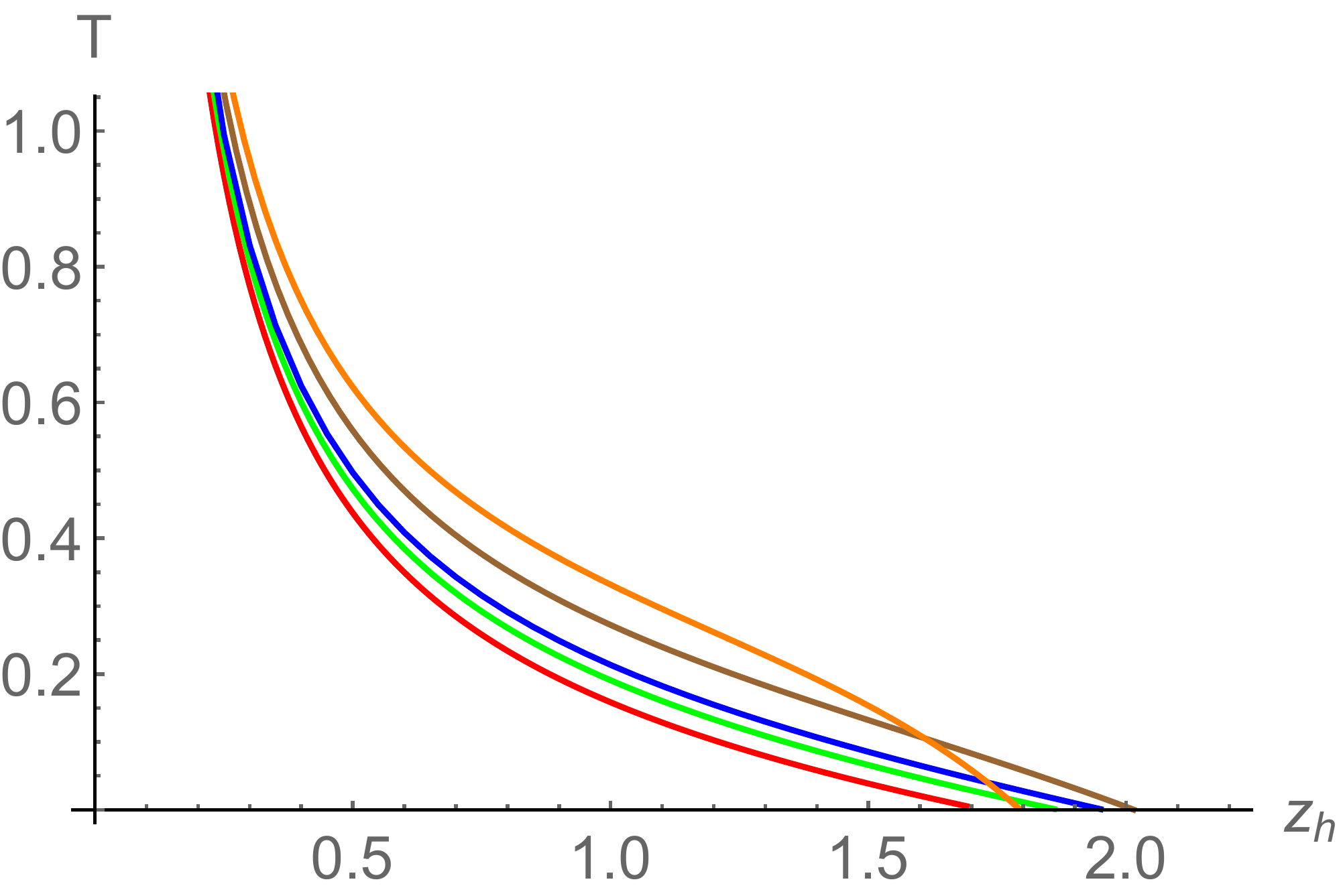}
\caption{ \small Hawking temperature $T$ as a function of horizon radius $z_h$ for various values of $a$.  Here $n=1$, $\mu=0.2$, $\kappa=-1$ and $D=4$ are used. Red, green, blue, brown and orange curves correspond to $a=0$, $0.3$, $0.5$, $1.0$ and $1.5$ respectively. }
\label{zhvsThvsamuPt2D4n1hyp}
\end{minipage}
\hspace{0.4cm}
\begin{minipage}[b]{0.5\linewidth}
\centering
\includegraphics[width=2.8in,height=2.3in]{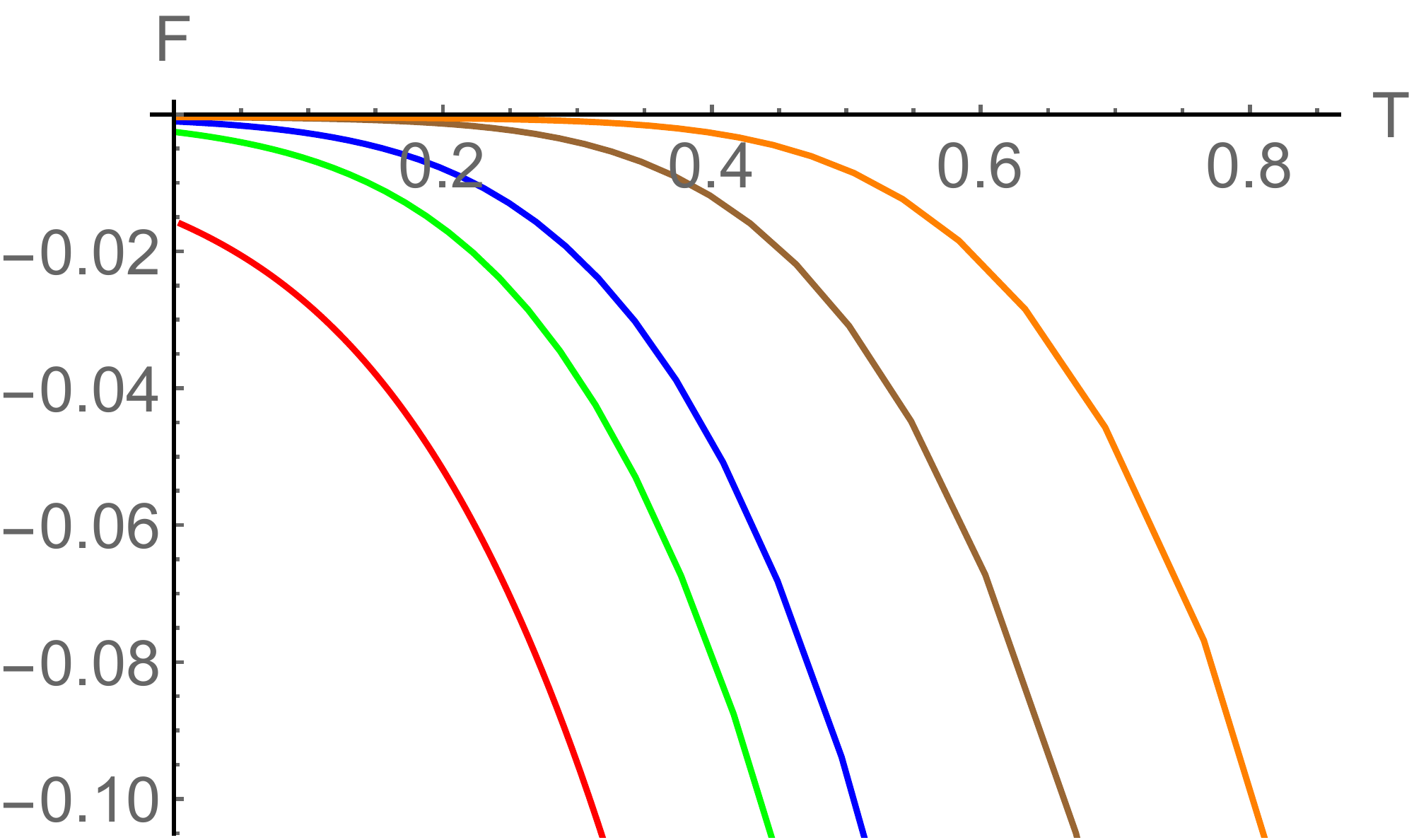}
\caption{\small Normalised Gibbs free energy $F$ as a function of Hawking temperature $T$ for various values of $a$. Here $n=1$, $\mu=0.2$, $\kappa=-1$ and $D=4$ are used. Red, green, blue, brown and orange curves correspond to $a=0$, $0.3$, $0.5$, $1.0$ and $1.5$ respectively. }
\label{TvsFvsamuPt2D4n1hyp}
\end{minipage}
\end{figure}

However, a few notable changes appear in the thermodynamic structure of the hyperbolic hairy black holes as compared to the planar and spherical hairy black holes. In particular, even though the specific heat of hairy hyperbolic black hole is always positive, suggesting it to be locally stable, however its free energy is always higher than the hyperbolic RN-AdS black hole. This suggests that, with the hyperbolic horizon,  the hairy black holes are always thermodynamically unfavored compared to RN-AdS at any temperature and there is no hairy/RN-AdS phase transition. The results are shown in Figures~\ref{zhvsThvsamuPt2D4n1hyp} and \ref{TvsFvsamuPt2D4n1hyp}. This result should be contrasted with the results of \cite{Gonzalez:2013aca}, where a phase transition to hairy configuration at low temperature was observed with the hyperbolic horizon.

\begin{figure}[h!]
\begin{minipage}[b]{0.5\linewidth}
\centering
\includegraphics[width=2.8in,height=2.3in]{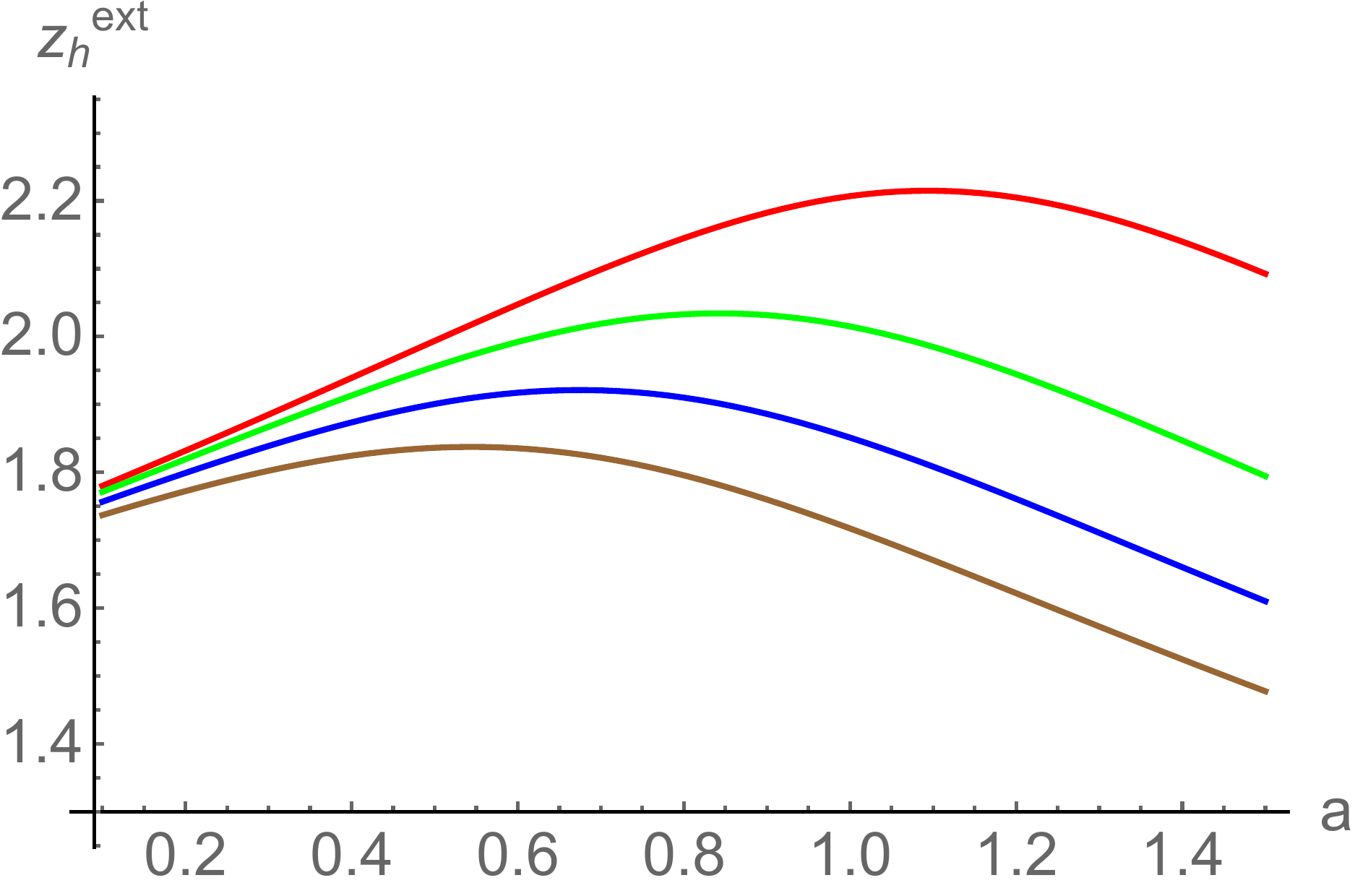}
\caption{ \small The variation of extremal black hole horizon radius $z_{h}^{\text{ext}}$ as function of $a$. Here $n=1$, $\kappa=-1$ and $D=4$ are used. Red, green, blue and brown curves correspond to $\mu=0.1$, $0.2$, $0.3$ and $0.4$ respectively.}
\label{avsZhextvsmuD4n1hyp}
\end{minipage}
\hspace{0.4cm}
\begin{minipage}[b]{0.5\linewidth}
\centering
\includegraphics[width=2.8in,height=2.3in]{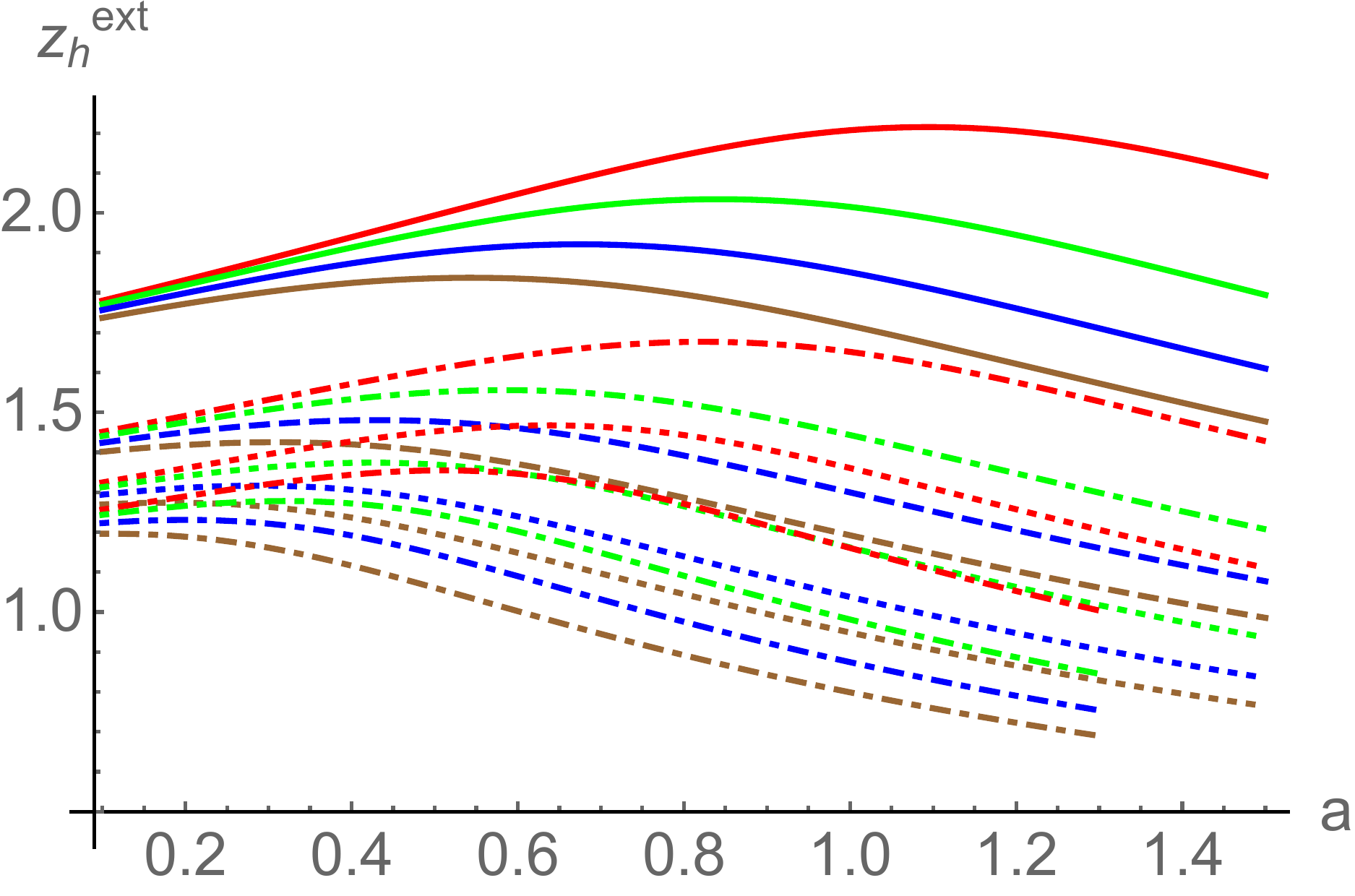}
\caption{\small The variation of extremal black hole horizon radius $z_{h}^{\text{ext}}$ as function of $a$. Here $n=1$ and $\kappa=-1$ are used. Red, green, blue and brown curves correspond to $\mu=0.1$, $0.2$, $0.3$ and $0.4$ respectively. Solid, dashed, dotted and dotdashed curves correspond to $D=4$, $5$, $6$ and $7$ respectively.}
\label{avsZhextvsmuvsDn1hyp}
\end{minipage}
\end{figure}

Moreover, the extremal horizon radius also shows distinct behaviour with the hyperbolic horizon. In particular, $z_{h}^{\text{ext}}$ shows non-monotonic behaviour with $a$. As $a$ increases, $z_{h}^{\text{ext}}$ first increases and then decreases. This is shown in Figure~\ref{avsZhextvsmuD4n1hyp}. The non-monotonic behaviour of $z_{h}^{\text{ext}}$ with the hyperbolic horizon should be contrasted with the planar and spherical horizons where $z_{h}^{\text{ext}}$ always decreases with $a$.

We further analyse the hyperbolic hairy black hole thermodynamics in higher dimensions and find similar results. In particular, the free energy of hyperbolic RN-AdS black hole is always smaller than the hyperbolic hairy black hole. Again, $z_{h}^{\text{ext}}$ depends non-monotonically on $a$. The only difference arises in the magnitude of $z_{h}^{\text{ext}}$ which decreases with $D$.  The dependence of $z_{h}^{\text{ext}}$ on $a$ and $\mu$ in different dimensions is shown in Figure~\ref{avsZhextvsmuvsDn1hyp}.

\section{Summary}
In this paper, we have considered $D$-dimensional gravity theory where the scalar field is minimally coupled to gravity along with a self-interacting potential and a $U(1)$ gauge field. We solved the coupled Einstein-Maxwell-Scalar equations of motion analytically and obtained an infinite family of exact charged hairy black hole solutions with a planar, spherical and hyperbolic horizon topologies in all the spacetime dimensions. The analytic solution is obtained in terms of a scale function $A(z)=-az^n$, which allowed us to introduce two parameters $a$ and $n$. The strength of the scalar hair is given by $a$, and in the limit $a\rightarrow0$ our solution reduces to the standard RN-AdS solution. We found that the scalar field is regular everywhere outside the horizon and goes to zero at the asymptotic AdS boundary. We calculated the Kretschmann scalar and found that the hairy black holes are devoid of any additional singularity outside the horizon. Similarly, the scalar potential is bounded from above from its UV boundary value and tends to the negative cosmological constant at the asymptotic boundary.

We then analysed the thermodynamic properties of the constructed hairy black hole solutions. We found that the charged planar hairy black holes have positive specific heat and are not only thermodynamically stable but also thermodynamically preferable. In particular, charged planar hairy black holes can have lower free energy than RN-AdS at low temperatures. This behaviour is in agreement with the holographic condensed matter systems. In these systems, a phase transition to hairy black hole configuration generally occurs below a critical temperature. This corresponds to the condensation of scalar field in the dual boundary theory. Our analysis further revealed that the temperature range for which the charged hairy black holes are thermodynamically preferable increases with the parameter $a$.  Similar results hold for the spherical charged hairy black hole. We did a thorough analysis of how parameters \{$a$, $\mu$, $n$, $\kappa$, $D$\} influence the hairy/RN-AdS critical temperature.  Unlike the charged case, the uncharged hairy black holes although have positive specific heat, however, always have higher free energy than their no-hair counterparts. On the other hand, hyperbolic black holes, both charged and uncharged, have higher free energy than RN-AdS at all temperature and, therefore, are thermodynamically unfavourable at all temperatures. We further investigated the dependence of the extremal horizon radius on $a$, $\mu$ and $D$.

There are many directions to extend our work. The first and foremost is to investigate the dynamical stability of the charged hairy black holes. In particular, it would be interesting to investigate the stability of the charged hairy black hole solutions against the scalar field perturbations. The quasinormal modes and time-domain analysis of the scalar perturbations will provide conclusive information about the dynamical stability of the hairy solution. Similarly, it would also be interesting to construct other charged hairy black hole solutions, in particular, by considering non-linear electrodynamic terms in gravity action. It would also be interesting to see whether the potential reconstruction method can be used to construct stable asymptotic flat hairy black hole backgrounds. Work in this direction is under progress.

\section*{Acknowledgments}
We would like to thank D. Choudhuri for careful reading of the manuscript and pointing out the necessary corrections. The work of SM is supported by the Department of Science and Technology, Government of India under the Grant Agreement number IFA17-PH207 (INSPIRE Faculty Award). The work of SP is supported by grant no.
16-6(DEC.2017)/2018(NET/CSIR) of UGC, India.

\end{document}